\documentclass{article}
\usepackage{ijcai25}

\usepackage{times}
\usepackage{soul}
\usepackage{url}
\usepackage[hidelinks]{hyperref}
\usepackage[utf8]{inputenc}
\usepackage[small]{caption}
\usepackage{graphicx}
\usepackage{amsmath}
\usepackage{amsthm}
\usepackage{booktabs}
\usepackage{algorithm}
\usepackage{algorithmic}
\usepackage[switch]{lineno}
\usepackage{tikz}
\usepackage{natbib}
\usetikzlibrary{decorations.pathreplacing}
\usepackage{enumerate}
\tikzset{
box/.style ={
circle,
minimum width =30pt,
minimum height =20pt, 
inner sep=4pt, 
draw=black,
fill = blue!20
}
}
\tikzset{
global scale/.style={
    scale=#1,
    every node/.append style={scale=#1}
  }
}
\usepackage{caption}
\usepackage{subcaption}
\usepackage{algorithm}
\usepackage{algorithmic}

\usepackage{amsmath}
\usepackage{amssymb}
\usepackage{multirow}
\usepackage{graphicx}
\usepackage{xcolor}
\usepackage{booktabs} 
\usepackage[most]{tcolorbox}
\usepackage{cleveref}
\usepackage{tablefootnote}
\usepackage{array}
\newcolumntype{K}[1]{>{\centering\arraybackslash}p{#1}}
\newcommand{\SW}{\operatorname{SW}}
\newcommand{\Rev}{\operatorname{Rev}}
\usepackage{cleveref}

\usepackage{xr}

\definecolor{lightblue}{RGB}{173, 216, 230} 
\definecolor{lightgreen}{RGB}{144, 238, 144} 
\definecolor{lightyellow}{RGB}{255, 255, 224} 
\definecolor{lightpink}{RGB}{255, 182, 193} 
\definecolor{lightgray}{RGB}{211, 211, 211} 
\definecolor{lightorange}{RGB}{255, 200, 160} 

\newtheorem{theorem}{Theorem}[section]
\newtheorem{definition}{Definition}[section]
\newtheorem{lemma}{Lemma}[section]
\newtheorem{proposition}{Proposition}[section]
\newtheorem{corollary}{Corollary}[section]
\newtheorem{remark}{Remark}

\title{Strategyproofness and Monotone Allocation of Auction in Social Networks}
\author{
Yuhang Guo$^{1,4}$
\and
Dong Hao$^{1,2}$\footnote{Corresponding Author.}\and
Bin Li$^{3}$\and
Mingyu Xiao$^1$\And 
Bakh Khoussainov$^1$\\
\affiliations
$^1$SCSE, University of Electronic Science and Technology of China\\
$^2$AI-HSS, University of Electronic Science and Technology of China\\
$^3$SCSE, Nanjing University of Science and Technology\\
$^4$ University of New South Wales
\emails
yuhang.guo2@unsw.edu.au, 
haodong@uestc.edu.cn,\\
cs.libin@njust.edu.cn, 
myxiao@uestc.edu.cn,
bmk@uestc.edu.cn
}

\begin{document}

\maketitle

\begin{abstract}
Strategyproofness in network auctions requires that bidders not only report their valuations truthfully, but also do their best to invite  neighbours from the social network.  In contrast to canonical auctions, where the value-monotone allocation in Myerson's Lemma is a cornerstone, a general principle of allocation rules for strategyproof network auctions is still missing. We show that, due to the absence of such a principle, even extensions to multi-unit network auctions with single-unit demand present unexpected difficulties, and all pioneering researches fail to be strategyproof.
For the first time in this field, we identify two categories of monotone allocation rules on networks: Invitation-Depressed Monotonicity (ID-MON) and Invitation-Promoted Monotonicity (IP-MON). They encompass all existing allocation rules of network auctions as specific instances. For any given ID-MON or IP-MON allocation rule, we characterize the existence and sufficient conditions for the strategyproof  payment rules, and show that among all such payment rules, the revenue-maximizing one exists and is computationally feasible. 
With these results, the  obstacle  of combinatorial network auction with single-minded bidders is now resolved.  
\end{abstract}

\section{Introduction}
In recent years,  auction  design in social networks has received emerging attention from the computer science and artificial intelligence community \citep{GUHA21a,LHG+22a}. In contrast to canonical auction theory, which concentrates solely on bidders directly reachable by the seller, network auction characterizes the auction  environment as large and unfixed, providing the potential to recruit additional participants. Existing works on network auctions mostly focus on devising mechanisms that motivate agents to actively disseminate auction information to their neighbors and invite their neighbours into the auction, thereby expanding the market, improving allocation efficiency, and simultaneously increasing the seller's revenue \citep{LHZ+17a,ZLX+18a,KBT+20a,LHG+22a}.

In classic auction theory, monotone allocation combined with critical value payment capture truthful mechanisms. For one-dimensional types, Myerson's lemma \citep{MYER81a} is the guiding principle. \citet{ARTA01a} further developed a concrete characterization of incentive-compatible single-parameter mechanisms based on Myerson's Lemma.
For multi-dimensional bidder types, incentive compatibility becomes more complex. For dominant strategy incentive-compatible (DSIC) and deterministic mechanisms, \citet{KEVI79a} proposed the monotonicity termed ``Positive Association of Differences'' (PAD) and showed that all the DSIC mechanisms are varieties of VCG mechanism \citep{VICK61a,CLAR71a,GROV73a} in unrestricted domain with at least three possible outcomes. \citet{ROCH87a} introduced cycle monotonicity in unrestricted domains, which is necessary and sufficient for DSIC. Weak-Monotonicity (W-MON), weaker than cycle monotonicity, was proposed in restricted domains \citep{LMN03a,BCL+06a}.   

For network auctions, the aforementioned monotonicity concepts are too general and impractical for strategyproofness. This is because the private types of bidders are shaped not only by their valuations but also by their social connections.  Agents must take into account invitation behaviors when formulating their bids, and vice versa. The mechanism   has to consider the complex preferences of agents over all combinations of bids and network structures, which seems implausible.  To date, no principle for allocation rules in strategyproof network auctions has been established. In sharp contrast to single-parameterized canonical auctions, where the value-monotonic allocation in Myerson's Lemma serves as a cornerstone, existing network auction mechanisms can only rely on loosely designed allocation rules based on trial and error.
Even extensions to multi-unit network auctions with single-unit demands pose unexpected difficulties. In the following sections, we show that the existing key mechanisms for multi-unit network auctions are not strategyproof.

For the theory of strategyproofness in network auctions, although \citet{LHZ20a} presented an elegant theorem to characterize all strategyproof network auctions, a characterization of \textit{monotone allocation in network auction} and the effect of monotone allocation on payment and strategyproofness is still missing, which has been a major obstacle for network auctions design. 
Currently, researchers are extending network auction designs to multi-unit scenarios \citep{ZLX+18a,KBT+20a,LLZ23a,FZL+23a}. Unfortunately, even seemingly straightforward extensions to multi-unit network auctions with \textit{single-unit demands} present unexpected difficulties, and all existing mechanisms turn out to be non-truthful or inefficient. For example, the first efforts of this line by \citet{ZLX+18a} and \citet{KBT+20a} are not strategyproof, allowing certain agents to gain by inviting fewer neighbours or even by not inviting at all. LDM \citep{LLZ23a} and MUDAN \citep{FZL+23a} use complex rules to localize bidders' competition to ensure strategyproofness, with significantly harmed efficiency and revenue as side effect. Please refer to \Cref{sec::related_work} for detailed related work.

\subsection{Our Contributions}
We study the theory of strategyproof network auctions, which provides a concise approach to achieving truthfulness and revenue optimization in network auctions. In particularly it is helpful for multi-unit or combinatorial network auctions, where most existing mechanisms fall short.

(1) We begin by examining why the Distance-based Network Auction with Multi-Unit (DNA-MU) mechanism \citep{KBT+20a} fails to achieve strategyproofness. We identify the underlying causes of this failure and propose a revised mechanism that restores strategyproofness. Our analysis reveals that merely enforcing value-monotonicity in allocation can complicate or even hinder the payment design.

(2) Given any value-monotone allocation, we identify a sufficient condition for the payment rule to be strategyproof in network auctions. We further characterize two categories of monotone allocation rules: Invitation-Depressed Monotonicity (\textbf{ID-MON}) and Invitation-Promoted Monotonicity (\textbf{IP-MON}). Each is grounded in different partial orderings in the bidders' multi-dimensional type space. Both ID-MON and IP-MON are not only value-monotone but also monotone with respect to network structure. Consequently, all existing strategyproof mechanisms with various allocation rules in network auctions can be explained by ID-MON or IP-MON.

(3) Building on ID-MON and IP-MON, we formally characterize the revenue-maximizing payment rules that satisfy individual rationality and strategyproofness. These payment rules establish the upper bound on the seller’s revenue achievable under any given ID-MON/IP-MON allocation rule.

(4) In sharp contrast to existing multi-unit network auctions, which are burdened by complex payment reasoning, our principles of ID-MON and IP-MON implementability greatly simplify the design of strategyproof network auction. To our knowledge, this is the first work to study a simple and principled framework for designing network combinatorial auction with single-minded bidders.

Results lacking full proofs are proven in the appendix.

\section{Preliminaries}
\subsection{Network Auction Model}
\begin{figure}[!htbp]
    \centering
\begin{tikzpicture}[global scale=0.75, box/.style={circle, draw}]
            \node[box,fill=gray,very thin](s) at(0,-0.5){{\LARGE$s$}};
            \node[box,fill={rgb:red,1;green,180;blue,80}](A) at(1.2,-1){$A$};
            \node[] at (1.7, -1){$4$};
            \node[box,fill=white](F) at(2.4,-1){$F$};
            \node[] at (2.9, -1){$6$};
            
            \node[box,fill={rgb:red,1;green,180;blue,80}](B) at(1.2,0){$B$};
            \node[] at (1.7, 0.3){$1$};
            \node[box,fill=white](C) at(2.4,0){$C$};
            \node[] at (2.9, 0.3){$4$};
            \node[box,fill=white](D) at(3.6,0){$D$};
            \node[] at (4.1, 0.3){$7$};
            \node[box,fill=white](H) at(4.8,0){$H$};
            \node[] at (5.3, 0.3){$5$};

            \draw[-,  line width=.8pt] (s) --(A);
            
            \draw[-,  line width=.8pt]  (s) --(B);
            \draw[-,  line width=.8pt]  (B) --(C);
            \draw[-,  line width=.8pt]  (C) --(D);
            \draw[-,  line width=.8pt]  (D) --(H);
            \draw[-,  line width=.8pt]  (B) --(F);

        \end{tikzpicture}
    \caption{Social network example with $6$ agents}
    \label{social-network-and-diffusion-example-1}
\end{figure}
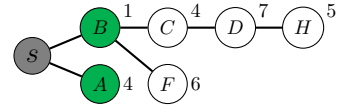

Consider a social network \(G=(N\cup\{s\},E)\), where \(N\cup\{s\}\) is the set of nodes while $E$ is the set of edges. $s$ is the seller node while agents in $N$ are potential bidders in the network. Denote each agent $i$'s neighbor set by $N(i)=\{j \mid (i,j) \in E\}$. Assume that the seller \(s\) has a collection of items \(\mathcal{K}\) (either homogeneous or heterogeneous) to be sold and initially she can only call together her direct neighbors $N(s)$ by herself. In order to expand the market, she can incentivize her neighbors to invite their own friends to join the market. \Cref{social-network-and-diffusion-example-1} is a social network $G=(N\cup\{s\},E)$ where $s$ is the seller and $N=\{A,B,C,F,D,H\}$ are potential bidders. 

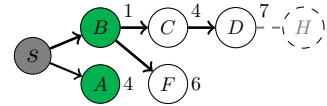
\begin{figure}[!htbp]
    \centering
\begin{tikzpicture}[global scale=0.75, box/.style={circle, draw}]
            \node[box,fill=gray,very thin](s) at(0,-0.5){{\LARGE$s$}};
            \node[box,fill={rgb:red,1;green,180;blue,80}](A) at(1.2,-1){$A$};
            \node[] at (1.7, -1){$4$};
            \node[box,fill=white](F) at(2.4,-1){$F$};
            \node[] at (2.9, -1){$6$};
            
            \node[box,fill={rgb:red,1;green,180;blue,80}](B) at(1.2,0){$B$};
            \node[] at (1.7, 0.3){$1$};
            \node[box,fill=white](C) at(2.4,0){$C$};
            \node[] at (2.9, 0.3){$4$};
            \node[box,fill=white](D) at(3.6,0){$D$};
            \node[] at (4.1, 0.3){$7$};
            \node[box,fill=white, dashed](H) at(4.8,0){\color{gray}{$H$}};
            
            \draw[->,  line width=.8pt] (s) --(A);
            
            \draw[->,  line width=1pt]  (s) --(B);
            \draw[->,  line width=1pt]  (B) --(C);
            \draw[->,  line width=1pt]  (C) --(D);
            \draw[-, dashed, gray, line width=.8pt]  (D) --(H);
            \draw[->,  line width=1pt]  (B) --(F);
        \end{tikzpicture}
    \caption{Information diffusion in the market}
    \label{social-network-and-diffusion-example-2}
\end{figure}

\Cref{social-network-and-diffusion-example-2} shows an instance of auction information diffusion. It starts from  seller $s$, who invites bidders $A$ and $B$ into the market. After that, bidder $B$ further invites  neighbors $C$ and $F$, then $C$ invites $D$. Later, bidder $D$ does not invite $H$, thus $H$ cannot enter the market. Finally, $\{A,B,C,D,F\}$ are the bidders. We denote $G=(N\cup\{s\}, E)$ as a digraph depicting the market with information diffusion.

Each bidder~$i$ has private information $\theta_i = (v_i, r_i)$, where $v_i$ represents her valuation, consistent with the classical single-parameter environment (e.g., single-item, $k$-unit with unit-demand, or knapsack auctions, etc), and $r_i = N(i)$ denotes the set of neighbors she can invite to participate in the auction.

Let $\mathbf{\theta}=(\theta_1,\ldots,\theta_n)$ be the type profile of the bidder set \(N\), and \(\mathbf{\theta}_{-i}=(\theta_1,\ldots,\theta_{i-1},\theta_{i+1},\ldots,\theta_n)\) be the type profile of the other bidders \(N\setminus\{i\}\). Define \(\Theta=\times_{i\in N} \Theta_i\) as the space of the joint type of bidders \(N\), where \(\Theta_i=\mathcal{R}_{\geq 0}\times \mathcal{P}(r_i)\) is the type space of bidder \(i\), and \(\mathcal{P}(r_i)\) is the power set of \(r_i\).  Each bidder \(i\) could strategically misreport. Let \(\theta_i^\prime=(v_i^\prime, r_i^\prime)\) be the reported type of bidder \(i\), where \(v_i^\prime \in \mathcal{R}_{\geq 0}\) and \(r_i^\prime \in \mathcal{P}(r_i)\).   

\begin{definition}
    A mechanism $\mathcal{M}=(f,p)$. Here $f$ is the allocation rule $f=(f_1,\cdots,f_n)$  and $p$ is the payment rule $p=(p_1,\cdots,p_n)$, where $f_i:\Theta \to \{0,1\}$ and $p_i:\Theta\to \mathcal{R}$. 
\end{definition}

Given any type profile \(\mathbf{\theta}^\prime\), we say \(f\) is \textit{feasible} if, for all \(\mathbf{\theta}^\prime \in \Theta\), it holds that the seller \(s\) can sell at most \(|\mathcal{K}|\) items. Let \(\mathcal{F}\) be the set of all feasible allocations.
For a given \(\mathbf{\theta}^\prime \in \Theta\) and a mechanism \(\mathcal{M} = (f, p)\),  the \textit{social welfare} is \(\SW(f, \mathbf{\theta}^\prime) = \sum_{i \in N} f_i(\mathbf{\theta}^\prime) v_i\). An allocation \(f^\ast \) is  \textit{efficient} if it always allocate goods to bidders who value them the most. Formally,

\begin{definition}[Efficiency (EF)]
    A network auction mechanism $\mathcal{M}=(f^\ast,p)$ is efficient if for all $\mathbf{\theta}^\prime \in \Theta$, $f^\ast = \arg\max_{f\in \mathcal{F}}\SW(f,\mathbf{\theta}^\prime).$
\end{definition}

For any \(\mathbf{\theta}^\prime\in \Theta\) and a mechanism \(\mathcal{M}=(f,p)\), the seller's \textit{revenue} is \(\Rev^{\mathcal{M}}(\mathbf{\theta}^\prime)=\sum_{i\in N}p_i(\mathbf{\theta}^\prime)\). Accordingly, we say that the mechanism \(\mathcal{M}\) is \textit{(weakly) budget balanced} if the seller never incurs negative revenue from the auction.

\begin{definition}[Weakly Budget Balance (WBB)]
    A network auction mechanism $\mathcal{M}=(f,p)$ is (weakly) budget balanced if $\forall \mathbf{\theta}^\prime \in \Theta, \Rev^{\mathcal{M}}(\theta^\prime)\geq 0$. 
\end{definition}

Given profile $\theta^\prime$, each bidder $i$'s \textit{quasi-linear} utility function is $u_i(\mathbf{\theta}^\prime, (f,p))=f_i(\theta^\prime)v_i - p_i(\theta^\prime)$. We simplify it as $u_i((v_i^\prime, r_i^\prime), \theta^\prime_{-i})$.
Next, we formulate  \textit{individual rationality} and \textit{strategyproofness}.

\begin{definition}[Individual Rationality (IR)]
    A network auction mechanism $\mathcal{M}=(f,p)$ is individual rational (IR) if for all $\mathbf{\theta}^\prime \in \Theta$, for all $i\in N,r_i^\prime\in \mathcal{P}(r_i)$, $u_i((v_i,r_i^\prime),\mathbf{\theta}^\prime_{-i})) \geq 0.$ 
\end{definition}
In network auction scenarios, given any mechanism $\mathcal{M}$, if for each bidder, truthfully disclosing her valuation brings non-negative utility, we call this mechanism satisfies IR. It does not place requirements on bidders' invitation behaviors.

\begin{definition}[Strategyproofness (SP)]
    A network auction mechanism $\mathcal{M}=(f,p)$ is strategyproof if for all $\mathbf{\theta}^\prime \in \Theta$, for all $ i\in N$, $u_i((v_i,r_i),\theta_{-i}^\prime) \geq u_i((v_i^\prime, r_i^\prime), \theta_{-i}^\prime).$ 
\end{definition}

In network auctions, strategyproofnes requires that, for any buyer, truthfully reporting her valuation and inviting \textbf{\textit{all}} the neighbors around is the dominant strategy.

\begin{definition}[Network-Implementable]\label{def_network_implementable}
    A network auction's allocation  $f$ is network-implementable if there exists a payment  $p$ such that $\mathcal{M}=(f,p)$ is strategyproof.
\end{definition}
Notably, network-implementable allocation  is different from implementable allocation in classic auctions. The mapping from allocation to payment takes  network structure (i.e., invitational incentive) into consideration. 

\begin{definition}[Degenerated]\label{degenerated_def}
    A network auction mechanism $\mathcal{M}=(f,p)$ is degenerated if for any profile $\theta \in \Theta$, for any bidder $i\in N$, and any invitation strategy $r_i^\prime \subseteq r_i$, $u_i((v_i,r_i^\prime), \theta^\prime_{-i}) = u_i((v_i,r_i), \theta^\prime_{-i})$.
\end{definition}

Intuitively, a network auction mechanism is degenerated if each agent's utility is independent of her invitation actions. 

\subsection{Strategyproof Network Auctions}
In classic auction theory, Myerson's Lemma \citep{MYER81a} presents the formulations for all individual rationality (IR) and strategyproof (SP) mechanisms under single-parameter domains. A normalized mechanism (where losers always pay zero) is considered SP \textit{if and only if} the allocation is value-monotone and winners always pay the critical winning bid.
\begin{definition}[Value-Monotonicity]
    Given a network auction mechanism $\mathcal{M}=(f,p)$ and profile $\theta^\prime$, for every bidder $i$, if allocation $f_i((v_i^\prime, r_i^\prime),\mathbf{\theta}^\prime_{-i})=1$ implies that $f_i((v_i^{\prime\prime},r_i^\prime),\mathbf{\theta}_{-i}^\prime)=1$ for any $v^{\prime\prime} \geq v^\prime$, then we say the allocation rule $f$ is value-monotone. 
\end{definition}
Value-monotonicity depicts that given any $r_i^\prime$ and $\theta_{-i}^\prime$, for any bidder $i$, increasing her bid $v_i^\prime$ will never turn herself from a winner to a loser. 
The characterization of IR and SP  in the context of network auctions was initially proposed by \citet{LHZ20a}. They established a sufficient and necessary conditions for IR and SP in network auctions. Before introducing the theorem, we first provide some essential definitions.

\begin{definition}[Payment Decomposition]
    Given a network auction mechanism $\mathcal{M}=(f,p)$ and any profile $\theta^\prime$ for any bidder $i$, her payment $p_i$ can be decomposed into the \textbf{winning payment} $\tilde{p}_i$ and the \textbf{losing payment} $\bar{p}_i$, such that $p_i(\theta^\prime) = f_i(\theta^\prime) \tilde{p}_i + (1-f_i(\theta^\prime))\bar{p}_i$. 
\end{definition}

\begin{definition}[Bid-Independent]
    Given a network auction mechanism $\mathcal{M}=(f,p)$ and profile $\theta^\prime$, $\forall ~ i \in N$, $v_i^\prime \neq v_i^{\prime\prime}$, $\tilde{p}_i((v_i^\prime,r_i^\prime),\theta_{-i})=\tilde{p}_i((v_i^{\prime\prime},r_i^\prime),\theta_{-i})$ and $\bar{p}_i((v_i^\prime,r_i^\prime),\theta_{-i})=\bar{p}_i((v_i^{\prime\prime},r_i^\prime),\theta_{-i})$.
\end{definition}

\begin{definition}[Invitational Monotonicity]\label{invitation-mono}
    Given a network auction mechanism $\mathcal{M}=(f,p)$, for each bidder $i$, fixing all other bidders' profile $\theta^\prime_{-i}$ and bid $v_i$, if her decomposed winning and losing payments satisfy that for all $r_i^\prime \subseteq r_i$, $
    \tilde{p}_i(v_i,r_i^\prime) \geq \tilde{p}_i(v_i,r_i) \, \wedge \, \bar{p}_i(v_i,r_i^\prime) \geq \bar{p}_i(v_i,r_i)$,
    then we say payment rule $p$ is invitational-monotone. 
\end{definition}
Intuitively, invitational-monotonicity of the payment represents that, regardless of being a winner or loser, when fixing a bidder's bid, inviting all the neighbors always minimizes their payment. This property directly incentivizes bidders to truthfully disclose their invitation sets.

\begin{definition}[Critical Winning Bid]\label{critical_winning_bid}
    Given $\mathcal{M}=(f,p)$, for any bidder $i$, fixing  others'  $\mathbf{\theta}^\prime_{-i}$, denote $v^\ast_i(r_i^\prime)$ as the critical winning bid for bidder $i$ when her invitation action is $r_i^\prime$:
\begin{equation}\label{eq_critical_winning_bid}
     v^\ast_i(r_i^\prime) = \inf_{v_i^\prime \in \mathcal{R}_{\geq 0}} \Big\{f_i((v_i^\prime, r_i^\prime), \mathbf{\theta}^\prime_{-i})=1\Big\}.  
\end{equation}
\end{definition}

The critical winning bid $v_i^\ast(r_i^\prime)$ is the minimum bid that makes bidder $i$ a winner, given all other bidders’ strategies $\mathbf{\theta}^\prime_{-i}$ and assuming bidder $i$ takes the invitation action $r_i^\prime$.

With the above definitions and decomposition, \citet{LHZ20a} proved a basic sufficient and necessary condition for all IR \& SP network auctions. 
\begin{theorem}[IR \& SP Network Auction \citep{LHZ20a}]\label{ic_diffusion_auction_theorem}
A network single-item auction mechanism \(\mathcal{M}=(f,p)\) is IR and SP if and only if, for all profiles \(\mathbf{\theta}\in \Theta\) and all bidders \(i\in N\), conditions (1)-(4) are satisfied:
\begin{enumerate}[(1)]
    \item The allocation rule \(f\) is value-monotone.
    \item \(\tilde{p}_i\) and \(\bar{p}_i\) are bid-independent and invitational-monotone.
    \item \(\tilde{p}_i(r_i)-\bar{p}_i(r_i)=v^\ast_i(r_i)\).
    \item \(\bar{p}_i(\emptyset)\leq 0\).
\end{enumerate}
\end{theorem}

Conditions (1) to (3) are for strategyproofness, while condition (4) is for IR.
However, \Cref{ic_diffusion_auction_theorem} only gives an abstract condition of the payment, it doesn't provide details regarding how to devise the allocation function $f$ or what the explicit form of the payment should be. To address this gap, a more fine-grained and operational characterization of strategyproofness is needed for network auction design. 

\section{$\mathcal{K}$-unit Network Auction}\label{section_3}
The Vickrey auction \citep{VICK61a} can be naturally extended to the multi-unit setting with unit-demand bidders in classical auction theory by allocating each unit to the top-$k$ highest bidders and charging each winner the $(k+1)$-st highest bid. However, this extension becomes substantially more complex in network auctions, which has sparked significant controversy and discussion in the community.

The very first work GIDM by \citep{ZLX+18a} tried to extend IDM \citep{LHZ+17a} into $k$-unit settings. However,  it is not strategyproof under some counterexamples constructed in \citep{TKT+19a}.
\citet{KBT+20a} proposed a new mechanism called DNA-MU to deal with \(k\)-unit network auctions with unit demand. Unfortunately, \citet{GHX+23a} proved that DNA-MU   also fails to be strategyproof for the same example in \citep{TKT+19a}. It is surprising that two totally different mechanisms fail to be strategyproof for the same counterexample. Upon encountering these difficulties, subsequent works on multi-unit network auctions either make strong assumptions about agents' information \citep{LLZ23a} or greatly compromise on their design objectives \citep{FZL+23a}. In this section, we will unveil the underlying reason of this failure and fix the DNA-MU mechanism.

\subsection{Counterexample of DNA-MU Mechanism}
We revisit DNA-MU in \Cref{DNA_MU_1}. It is based on a key notion, invitational-domination, which is widely used in network auctions.

\begin{definition}[Invitational-Domination]
    Given a digraph \(G = (N \cup \{s\}, E)\), for any two bidders \(i, j \in N\), \(i\) invitationally-dominates \(j\) if and only if all the paths from seller \(s\) to \(j\) must include \(i\).
\end{definition}

Intuitively, if bidder \(A\) dominates \(B\), then without \(A\)'s invitation, it is impossible for \(B\) to enter the auction market. Merging the invitational-domination relations between every pair of nodes, one can create the invitational-domination tree (IDT) for all informed bidders in $G$. IDT is  a partial ordering of agents regarding their topological importance.

Based on these concepts, we introduce the DNA-MU mechanism, which initially determines a distance-based priority ordering using the classic Breadth-First Search (\texttt{BFS}) algorithm. In this priority ordering, it has been proved that no bidder can improve her priority by misreporting her type. Next, it decides whether to allocate one item to each bidder via a threshold bid $v^k(N\setminus (T_i\cup W))$ \footnote{$v^k(N\setminus (T_i\cup W))$ is the $k$-th  bid in bidder set $N\setminus (T_i\cup W)$.} where $k$ is dynamically updated, $T_i$ is the sub-tree rooted at $i$ in the IDT, containing all the bidders who are invitationally dominated by $i$, and $W$ is the winner set. See \Cref{DNA_MU_1} for details of DNA-MU.

\begin{algorithm}[!htbp]
\caption{DNA-MU Mechanism \citep{KBT+20a}}
\label{DNA_MU_1}
\begin{algorithmic}[1]
\REQUIRE $G=(N\cup\{s\}, E)$, $\mathbf{\theta}$, $\mathcal{K}$;\\
\ENSURE Allocation $f$, payment $p$;
\STATE Initialize ordering $\mathcal{O}\leftarrow \mathtt{BFS}(G, s)$;\\
\STATE Create Invitational-Domination Tree (IDT) $T$;\\
\STATE Initialize $k \leftarrow |\mathcal{K}|, W\leftarrow \emptyset$;\\
\FOR{$i$ in $\mathcal{O}$}
\STATE $T_i\leftarrow $ Sub-Tree rooted by $i$ in $T$;\\
\IF{{\color{blue}{$v_i \geq v^k(N\setminus (T_i\cup W))$}}}
\STATE $f_i\leftarrow 1,{\color{blue}{p_i \leftarrow v^k(N\setminus (T_i\cup W))}}$;\\
\STATE Update ${\color{blue}{k \leftarrow k - 1}}, W\leftarrow W\cup \{i\}$;\\
\ENDIF
\ENDFOR
\STATE \textbf{Return} $f, p$.
\end{algorithmic}
\end{algorithm}

However, DNA-MU fails to be strategyproof in cases where some losers can invite fewer neighbors to gain an extra benefit. We run DNA-MU for the social network in \Cref{social-network-and-diffusion-example-1} with two  invitation profiles: one where all the bidders fully invite their neighbors (truthful behavior) and the other where bidder \(D\) does not invite \(H\) (false behavior), as shown in \Cref{social-network-and-diffusion-example-2}. \Cref{dna_mu_table1} shows that bidder \(D\) is profitable by exhibiting false behavior, making DNA-MU fail to be strategyproof.

\begin{table}[!htbp]
    \centering
\caption{Results of DNA-MU in \Cref{social-network-and-diffusion-example-1} with different $r_D$}
\scalebox{0.75}{
\begin{tabular}{@{}K{2cm}K{2cm}K{5.5cm}@{}}
\toprule
       &Allocation   & Payment  \\ 
\midrule
{$r_D=\{H\}$} & {$\{B,F,C\}$} & {$A(0),B(0),F(5),C(4),\bf{D(0)},H(0)$} \\
\midrule
{$r_D^\prime=\emptyset$} & {$\{A,B,\bf{D}\}$} & {$A(4),B(0),F(0),C(0),\bf{D(6)},H(0)$} \\ 
\bottomrule 
\end{tabular}
}
\label{dna_mu_table1}
\end{table}

\begin{proposition}\label{theorem_dna_mu_fail_ic}
    In $\mathcal{K}$-unit network auctions, DNA-MU mechanism is not SP when $|\mathcal{K}|\geq 3$.
\end{proposition}
Please refer to the appendix for more detailed running procedures and a general proof of \Cref{theorem_dna_mu_fail_ic}.

\subsection{Reason and a Correction}

The direct reason why DNA-MU fails to be strategyproof is that its payment rule is not invitationally-monotone, contradicting condition (2) in \Cref{ic_diffusion_auction_theorem}. This is easily proved in the following \Cref{DNA_MU_payment_NOT_referral_monotone}.

\begin{proposition}\label{DNA_MU_payment_NOT_referral_monotone}
    The payment rule  of mechanism DNA-MU is not invitationally-monotone.
\end{proposition}
\begin{proof}
Consider the counterexample in \Cref{social-network-and-diffusion-example-2}.
Bidder $D$ has two possible invitation strategies $r_D^1=\emptyset$ and $r_D^2=\{H\}$. Notice that  $\tilde{p}_D(r_D^1)=6$ while $\tilde{p}_{D}(r_D^2)=v^\ast_D(r_D^2)=+\infty > \tilde{p}_D(r_D^1)$. However, invitational-monotonicity (\Cref{invitation-mono}) requires 
$\tilde{p}_D(r_D^2)\leq \tilde{p}_D(r_D^1)$ as $r_D^1\subseteq r_D^2$, implying DNA-MU payment fails invitational monotonicity.
\end{proof}

We revise both the allocation and payment rules of DNA-MU to design a new mechanism, termed DNA-MU-Refined (DNA-MU-R), which restores strategyproofness. Specifically, in line 8, we adjust the threshold condition from {\color{blue}{$v_i \geq v^k(N \setminus (T_i \cup W))$}} to {\color{red}{$v_i \geq v^k(N \setminus T_i)$}}. In lines 9-10, we update the payment rule from {\color{blue}{$p_i \leftarrow v^k(N \setminus (T_i \cup W))$}} to {\color{red}{$p_i \leftarrow v_i^\ast(r_i)$}} and remove {\color{red}{the decrement of the parameter}} {\color{blue}{$k$ ($k\leftarrow k - 1$)}}. The formal algorithm for the DNA-MU-R mechanism is deferred to the appendix.

\begin{lemma}\label{DNA_MU_R_properties}
    DNA-MU-R Mechanism is IR, SP, and WBB. 
\end{lemma}

\Cref{dna_mu_r_table1} presents the results of DNA-MU-R in the counterexample. We defer the proof of \Cref{DNA_MU_R_properties} and the detailed execution steps of DNA-MU-R to the appendix.

\begin{table}[!htbp]
    \centering
    \caption{Results of DNA-MU-R in \Cref{social-network-and-diffusion-example-1} with different $r_D$}
\scalebox{0.75}{
\begin{tabular}{@{}K{2cm}K{2cm}K{5.5cm}@{}}
\toprule
       &Allocation   & Payment  \\ 
\midrule
{$r_D=\{H\}$} & {$\{B,F,C\}$} & $A(0),B(0),F(4),C(1),{\bf{D(0)}},H(0)$ \\
\midrule
{$r_D^\prime=\emptyset$} & {$\{A, B,F\}$} & $A(4),B(0), F(4),C(0), {\bf{D(0)}},H(0)$ \\ 
\bottomrule 
\end{tabular}
}
\label{dna_mu_r_table1}
\end{table}

Regarding DNA-MU-R mechanism,
we have revised both the allocation (line 8 and line 10) and the payment rules (line 9 and line 10). Is is possible to only revise the payment (line 9)? The answer is unknown yet.
Technically, the losing payment $\bar{p}_i(r_i)$ is always zero. 
Thus to satisfy condition (3)  in \Cref{ic_diffusion_auction_theorem}, the winning payment $\tilde{p}_i(r_i)$ must be equal to the critical winning bid $v^\ast_i(r_i)$.
However, the allocation rule of DNA-MU leads to a bad consequence that, bidder $D$'s critical winning bid, which is determined by allocation, $v^\ast_D(r_D)$ is not invitationally-monotone. It is worth noting that GIDM  \citep{ZLX+18a} also fails to be invitationally-monotone. 

\begin{remark}
{\textit{Merely ensuring value-monotonicity} for the allocation can \textit{complicate or even fail} the  payment design. }
\end{remark}

Then what kind of allocation rules in network auctions can ensure network-implementability? In the following section, we will reveal, from a high-level perspective, the characteristics of allocation rules that are considered ``good'' in the context of network auction design.

\section{Monotonicity and Implementability}\label{Section_4}

In classic auction theory, value-monotone allocation with critical payment play the vital role for strategyproofness. In this section, we first propose general payment functions satisfying strategyproofness for any given value-monotone allocation rule, based on the fundamental principles in \Cref{ic_diffusion_auction_theorem}. We then identify two  classes of network-implementable allocation rules: Invitation-Depressed Monotonicity (ID-MON) and Invitation-Promoted Monotonicity (IP-MON). 

Technically, we will unveil the fundamental reason why these seemingly opposite allocations can both be implemented in strategyproof network auctions, despite their significantly different performances. Furthermore, we derive the revenue-maximizing payment rule for both ID-MON and IP-MON allocations. Specifically, given any ID-MON or IP-MON allocation, our payment schemes achieve the upper bound of the seller's revenue within the strategyproof domain.

\subsection{Value-Monotone Allocation}
Based on \Cref{ic_diffusion_auction_theorem}, given any value-monotone allocation $f$, we characterize a general class of payment schemes $p$ such that if such a $p$ exists, then  $f$ is \emph{network-implementable}.

\begin{corollary}\label{ic_payment_scheme}
    Given any value-monotone allocation $f$ and profile $\theta$, for each bidder $i$ with $\theta_i=(v_i,r_i)$, if there exists some payment function $\tilde{p}(r_i)=\tilde{g}(r_i) + h(\theta_{-i})$ and $\bar{p}(r_i)=\bar{g}(r_i) + h(\theta_{-i})$ such that the following conditions hold for functions $\tilde{g}(\cdot)$, $\bar{g}(\cdot)$,
    \begin{equation*}
    \begin{aligned}
    \tilde{g}(r_i) - \bar{g}(r_i) &= v^\ast(r_i), \\
    r_i = \arg\min_{r_i^\prime\subseteq r_i} \tilde{g}(r_i^\prime),& \, r_i = \arg\min_{r_i^\prime \subseteq r_i} \bar{g}(r_i^\prime),
    \end{aligned}
    \end{equation*}
    then  $f$ is network-implementable.
\end{corollary}

\Cref{ic_payment_scheme} provides a basic guideline for designing strategyproof payments when   a value-monotone allocation function is given. 
The critical winning bid $v_i^\ast(r_i)$ in the first constraint is determined by the allocation rule. This raises an immediate question: can we leverage the monotonicity over \( r_i \) to develop  network-implementable allocation rules?
 
Another significant observation is that given any value-monotone allocation, when fixing all other bidders' reporting type, the critical winning bid for each bidder $i$ across two different invitation strategies $r_i^1,r_i^2 \subseteq r_i$ is comparable. 

\begin{lemma}\label{critial_bid_with_allocation}
    Given any value-monotone allocation rule $f$, for any bidder $i$, fixing all other bidders' profile $\theta_{-i}$, for two different invitation strategies $r_i^1 \subseteq r_i,r_i^2 \subseteq r_i$, $v^\ast_i(r_i^1) \leq v^\ast_i(r_i^2)$ if and only if $\forall\, v_i\in \mathcal{R}_{\geq 0}$, $f(v_i,r_i^1) \geq f(v_i,r_i^2)$. 
\end{lemma}

Next, we specify two different types of value-monotone allocation functions: Invitational-Depressed Monotonicity (ID-MON) and Invitational-Promoted Monotonicity (IP-MON). With these allocation functions, a strategyproof payment scheme always exists, meaning that the ID-MON and IP-MON allocation rules are always network-implementable.

\subsection{Invitation-Depressed Monotone Allocation} \label{subsection3-1}

Invitation-Depressed Monotone (ID-MON) allocation was initially defined in \citep{LHZ20a}. ID-MON allocations are based on the economic intuition that invitations can attract more bidders, thereby intensifying competition in the market and making it harder for each bidder to win. Although ID-MON favors bidders who invite fewer neighbors, it does not inherently incentivize invitations. Therefore, during the payment design, the auctioneer should compensate the bidders to encourage more invitations.

Technically, ID-MON allocation is based on a partial ordering $\succeq_{\mathcal{D}}$ over bidders' type profile  $\theta$. 
\begin{definition}
  For any bidder $i$ and  two  types $\theta_i^1=(v_i^1,r_i^1)$  and $\theta_i^2=(v_i^2,r_i^2)$, denote the invitation-depressed partial order  by $\succeq_{\mathcal{D}}$: if $v_i^1\geq v_i^2 $ and $ r_i^1\subseteq r_i^2$,  then $\theta_i^1 \succeq_{\mathcal{D}} \theta_i^2$.
\end{definition}
By leveraging this invitational-depressed partial ordering, the definition of ID-MON is as follows.

\begin{definition}[Invitation-Depressed Monotonicity (ID-MON)] 
Given an allocation rule $f$ and all other bidders' profile $\theta_{-i}$, If, for every bidder \(i\), the allocation \(f_i(\theta_i, \mathbf{\theta}_{-i}) = 1\) implies that for all \(\theta_i^\prime \succeq_{\mathcal{D}} \theta_i\), \(f_i(\theta_i^\prime, \mathbf{\theta}_{-i}) = 1\), then we say the allocation \(f\) is invitation-depressed monotone.
\end{definition}

\begin{lemma} \label{critical_bid_monotonicity_competitive}
Given an ID-MON allocation \( f \), for each bidder \( i \in N \) and   two type profiles \( \theta_i^1 = (v_i, r_i^1) \) and \( \theta_i^2 = (v_i, r_i^2) \), where \( r_i^1 \subseteq r_i^2 \), it always holds that \( v_i^\ast(r_i^1) \leq v_i^\ast(r_i^2) \).
\end{lemma}

The following theorem shows that for any ID-MON allocation rule \( f \), there always exists a payment scheme \( p \) such that \((f, p)\) is strategyproof.
\begin{theorem}\label{ic_implementability_id_mon}
    Every ID-MON allocation $f$ is network-implementable. 
\end{theorem}

With regard to the implementation of a strategyproof mechanism, by leveraging the features of ID-MON, we present the following theorem to demonstrate how to construct the optimal payment \( p^\ast \) such that \((f, p^\ast)\) is strategyproof and maximizes the seller's revenue. Please refer to the appendix for detail proof of \Cref{opt_payment_id_monotonicity}.
 
\begin{theorem}\label{opt_payment_id_monotonicity}
    Given an ID-MON allocation $f$ and profile $\theta$, for each bidder $i$, let $\tilde{p}_i=v_i^\ast(\emptyset)$,  $\bar{p}_i=v_i^\ast(\emptyset) - v_i^\ast(r_i)$, and $p^\ast=\{f_i(\theta)\tilde{p}_i + (1-f_i(\theta))\bar{p}_i\}_{i\in N}$, then $\mathcal{M}^\ast=(f,p^\ast)$ is IR and SP, and for any other IR and SP  $\mathcal{M}^\prime =(f,p^\prime)$, $\Rev^{\mathcal{M}^\ast}(\theta) \geq \Rev^{\mathcal{M}^\prime}(\theta)$.
\end{theorem}

It is not hard to see that the social welfare maximizing (efficient) allocation rule satisfies ID-MON. Therefore, although directly extending  VCG  into network  does not maximize the seller's revenue \citep{LHZ+17a,LHG+22a}, we can obtain the maximum revenue by utilizing the payment  in \Cref{opt_payment_id_monotonicity}:
\begin{enumerate}
    \item Allocate items to bidders with top-$k$ highest bids.
    \item Create the Invitational-Domination Tree (IDT) $T$.
    \item Each winner $i$ pays $v^k(N\setminus T_i)$ while each loser $j$ ``pays" $v^k(N\setminus T_j) - v^k(N)$.
\end{enumerate}
Call this mechanism VCG-Revenue-Maximizing (VCG-RM) mechanism. We can obtain the following statement.
\begin{corollary}\label{VCG_RM_properties}
    In $k$-unit network auction with single-unit demand bidders, given profile $\theta$, VCG-RM mechanism is EF, IR, SP, and $\Rev^{\text{VCG-RM}}(\theta) \geq \Rev^{\text{VCG}}(\theta)$.
\end{corollary}

\subsection{Invitation-Promoted Monotone Allocation}\label{subsection3-2}
In contrast to ID-MON allocations, we now characterize another class of monotone allocations that allocate items in the opposite manner. The intuition is rather straightforward: the more neighbors that bidders introduce into the auction, the higher their contribution. Therefore, the allocation rule should favor bidders who have more neighbors. We refer to this class of allocation as Invitation-Promoted Monotone (IP-MON) allocation. They are based on an opposite partial ordering of bidders’ types.

\begin{definition}\label{ip_partial_ordering}
  For any bidder $i$ and  two  types $\theta_i^1=(v_i^1,r_i^1)$  and $\theta_i^2=(v_i^2,r_i^2)$, denote the invitation-promoted partial ordering by $\succeq_{\mathcal{P}}$: if $v_i^1\geq v_i^2 $ and $ r_i^2\subseteq r_i^1$,  then $\theta_i^1 \succeq_{\mathcal{D}} \theta_i^2$.
\end{definition}

 The difference from this partial ordering with that for ID-MON is the ordering on $r_i$ is totally converse. Given partial order $\succeq_{\mathcal{P}}$, we propose the following monotonicity.

\begin{definition}[Invitation-Promoted Monotonicity (IP-MON)]
    Given an allocation rule $f$ and all other bidders' profile $\theta_{-i}$, if for every bidder $i$, the allocation $f_i(\theta_i,\mathbf{\theta}_{-i})=1$ implies that for all $\theta_i^\prime \succeq_{\mathcal{P}} \theta_i, f_i(\theta_i^\prime, \mathbf{\theta}_{-i})=1$, then we say allocation $f$ is invitation-promoted monotone.
\end{definition}

Similarly, we introduce the monotonicity of critical winning bid for IP-MON allocation rules.
\begin{lemma}\label{critial_bid_IP_monotonicity}
    Given an IP-MON allocation $f$, for each bidder $i\in N$, for any two type profile $\theta_i^1=(v_i, r_i^1)$ and $\theta_i^2(v_i,r_i^2)$ where $r_i^2\subseteq r_i^1$, it always holds that $v_i^\ast(r_i^1) \leq v_i^\ast(r_i^2)$.
\end{lemma}

\begin{theorem}\label{ic_implementability_ip_mon}
    Every IP-MON allocation rule $f$ is network-implementable.
\end{theorem}

Following the idea in ID-MON subsection, we can derive the revenue-maximizing payment for any IP-MON allocations. See the appendix for detailed proof of \Cref{opt_payment_IP_monotonicity}.
\begin{theorem}\label{opt_payment_IP_monotonicity}
    Given an IP-MON allocation  $f$ and profile $\theta$, for each bidder $i$, let $\tilde{p}_i=v_i^\ast(r_i)$,  $\bar{p}_i=0$, and $p^\ast=\{f_i(\theta)\tilde{p}_i + (1-f_i(\theta))\bar{p}_i\}_{i\in N}$, then $\mathcal{M}^\ast=(f,p^\ast)$ is IR and SP, and for any other IR and SP mechanism $\mathcal{M}^\prime=(f,p^\prime)$, $\Rev^{\mathcal{M}^\ast}(\theta) \geq \Rev^{\mathcal{M}^\prime}(\theta)$.
\end{theorem}

It is worth noting that the DNA-MU-Refined mechanism we proposed in the last section satisfy IP-MON and the payment rule has maximized the seller's revenue.  
\begin{proposition}\label{DNA_MU_R_allocation_IP_MON}
The allocation  of DNA-MU-R  is IP-MON.
\end{proposition}

\subsection{Insights and Implementation Complexity}
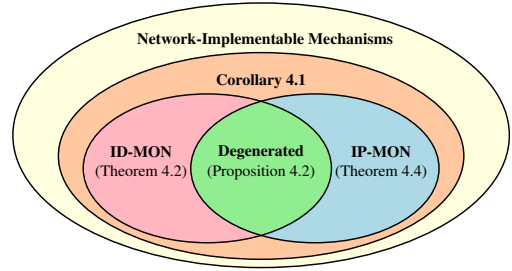
\begin{figure}[!htbp]
    \centering
    \scalebox{0.55}{
\begin{tikzpicture}
    \fill[lightyellow] (0, 0.5) ellipse (6 and 3.2);
    \node at(0.1, 2.8) {\large \textbf{Network-Implementable Mechanisms}};
    \fill[lightorange] (0,0) ellipse (4.9 and 2.5);
    \node at(0, 1.8) {\large \textbf{\Cref{ic_payment_scheme}}};
    \fill[lightpink] (-1.3,-0.3) ellipse (3 and 1.8);
    \node at (-2.9,  0.1) {\large \textbf{ID-MON}};
    \node at (-2.9, -0.4) {\large (\Cref{opt_payment_id_monotonicity})};
    \fill[lightblue] (1.3,-0.3) ellipse (3 and 1.8);
    \node at (2.9,  0.1) {\large \textbf{IP-MON}};
    \node at (2.9, -0.4) {\large (\Cref{opt_payment_IP_monotonicity})};
    
    \begin{scope}
        \clip  (-1.3,-0.3) ellipse (3 and 1.8);
        \fill[lightgreen]  (1.3,-0.3) ellipse (3 and 1.8);
    \end{scope}
    \node at (0,  0.1) {\large \textbf{Degenerated}};
    \node at (0, -0.4) {\large (\Cref{degenerate_lemma})};
    \draw[line width=1pt] (0, 0.5) ellipse (6 and 3.2);
    \draw[line width=1pt] (0,0) ellipse (4.9 and 2.5);
    \draw[line width=1pt] (-1.3,-0.3) ellipse (3 and 1.8);
    \draw[line width=1pt]  (1.3,-0.3) ellipse (3 and 1.8);
\end{tikzpicture}}
    \caption{Relations of mechanisms identified in this paper.}
    \label{truthful_framework_pic}
\end{figure}

The mechanism design space in the network auction scenario is extremely large. Given one  value-monotone allocation rule, there could be various payment rules that make the mechanism strategyproof. We summarize our results in the previous subsections in \Cref{truthful_framework_pic}, which shows the relations between each category of strategyproof mechanisms.
 
\noindent (\colorbox{lightyellow}{\textbf{Yellow}}) In \Cref{ic_diffusion_auction_theorem}, value-monotonicity is  sufficient and necessary  to guarantee truthfulness.\\
 (\colorbox{lightorange}{\textbf{Orange}}) In \Cref{ic_payment_scheme}, functions $\tilde{g}(\cdot)$ and $\bar{g}(\cdot)$ are introduced to specify SP network auctions.\\
(\colorbox{lightpink}{\textbf{Pink}}/\colorbox{lightblue}{\textbf{Blue}}) Since bidders' type is two-dimensional, only considering value-monotone  allocation while ignoring the monotonicity in the invitation dimension makes devising the SP payment, i.e., the $\tilde{g}(\cdot)$ and $\bar{g}(\cdot)$ functions be much more complicated. Therefore, we identify  ID-MON and IP-MON and proved both of these two  monotone allocations can be sufficient to achieve strategyproofness.\\
(\colorbox{lightgreen}{\textbf{Green}}) IP-MON and ID-MON  can be considered as two special classes within the design paradigm in \Cref{critial_bid_with_allocation}. By instantiating the partial order in the invitation \( r_i \) dimension,   \(\tilde{g}(\cdot)\) and \(\bar{g}(\cdot)\) are constructed. There exists a class of strategyproof mechanisms at the intersection of these two classes, which is degenerated in the sense of \Cref{degenerated_def}.

\begin{proposition}\label{degenerate_lemma}
    Given any mechanism $\mathcal{M}=(f,p)$ where $f$ satisfies both ID-MON and IP-MON, and $p$ satisfies the revenue-maximizing payment schemes in \Cref{opt_payment_id_monotonicity} and \Cref{opt_payment_IP_monotonicity}. Then $\mathcal{M}$ is degenerated.
\end{proposition}
     
\begin{corollary}\label{special_cases_of_lemma_2}
    ID-MON and IP-MON allocations with payment scheme in \Cref{opt_payment_id_monotonicity} and \Cref{opt_payment_IP_monotonicity} are special cases in \Cref{ic_payment_scheme}.
\end{corollary}

The ideas of SP auction design for these two monotonicities are entirely different. ID-MON-based mechanisms should follow the principle that \textit{losers could get potential benefits on payoffs} to incentivize more invitations, even though ID-MON allocation itself discourages invitations. On the other hand, IP-MON-based mechanisms are more straightforward in incentivizing invitations since \textit{the allocation itself already guarantees that}. These two allocation rules achieve distinct trade-offs between social welfare and revenue.

\begin{proposition}\label{EFF_WBB_2_MONO}
    Under IR and SP constraint, there exists an instance such that no IP-MON allocation is EF. Mechanism under ID-MON may fail WBB.
\end{proposition}

Another positive result is that both ID-MON and IP-MON mechanisms are computationally feasible.

\begin{proposition}\label{ID_IP_MON_computational_tractable}
If an ID-MON or IP-MON allocation $f$ runs in polynomial time $O(T)$, then the revenue-maximizing payment $p^\ast$ is computed in $O(N\cdot T \log (\max_{i\in N}v_i))$.
\end{proposition}

Furthermore, it is interesting to note that the allocation rules in all existing network auction mechanisms are either ID-MON, IP-MON, or fall within their intersection.
We categorize the existing mechanisms into three groups, as presented in \Cref{classification_mechanism_monotonicity_table} (\Cref{classification_existing_sp_mechanisms}). Based on the above analysis, especially \Cref{ic_implementability_id_mon} to \Cref{EFF_WBB_2_MONO} and \Cref{ID_IP_MON_computational_tractable}, we emphasize the following major result.

\begin{remark}\label{remark_2}
{Designing strategyproof network auctions can boil down to \textit{finding ID-MON and IP-MON allocations and applying the corresponding revenue-maximizing payments} in \Cref{opt_payment_id_monotonicity} and \Cref{opt_payment_IP_monotonicity}. }  
\end{remark}

The above characterization can guide the design of strategyproof mechanisms for network auctions with complex tasks. Given an ID-MON or IP-MON allocation, we can design an appropriate payment in a simple manner. For example, guided by \Cref{opt_payment_id_monotonicity}, we easily proposed revisions to VCG and DNA-MU, which are called VCG-RM and DNA-MU-R, respectively. We showcase that the VCG-RM and DNA-MU-R mechanisms significantly outperform the only two existing strategyproof mechanisms LDM-Tree \citep{LLZ23a} and MUDAN \citep{FZL+23a} regarding  social welfare and revenue in the appendix.

\section{Combinatorial Network Auction}\label{section_5}

Regarding combinatorial network auctions with single-minded bidders, \citet{FZL+24a} recently proposed the LOS-SN mechanism, inspired by the MUDAN mechanism \citep{FZL+23a}, introducing a novel approach to establishing the priority order. Notably, we observe that the LOS-SN mechanism satisfies the IP-MON condition, bringing it within the framework of \Cref{ic_implementability_ip_mon}. Building on this, we revisit the problem by naturally extending the classic results from combinatorial auctions with single-minded bidders to network auction settings, addressing both ID-MON and IP-MON allocations in a more accessible manner. 

 The scenario is specified as follows: seller \( s \) possesses a set \(\mathcal{K}\) of \( k \) heterogeneous items. For each bidder \( i \in N \), she is  single-minded if and only if there exists a unique bundle of goods \( S^\ast_i \subseteq \mathcal{K} \) that bidder \( i \) favors. Formally, each bidder's valuation can be represented by \( v_i(S) = v_i \) if and only if \( S = S^\ast_i \), and for all other bundles \( S^\prime \neq S \), \( v(S^\prime) = 0 \).
For all bidders, their favorite bundles are public information, while their private information is two-dimensional: the bid \( v_i \) for \( S^\ast_i \) and the invitation set \( r_i \). It is well-established that finding an efficient allocation for this problem is NP-hard \citep{LOL+02a}. This complexity extends to network auctions, where each classical scenario can be interpreted as a networked case in which all the bidders are directly connected to the seller. Furthermore, it has been shown that there exists no polynomial time algorithm for optimal allocation with an approximation ratio better than \( k^{1/2-\epsilon} \). The well-known near-optimal approximation scheme is presented in \Cref{sqrt_k_approximation} \citep{LOL+02a}.

 \begin{algorithm}[!htbp]
\caption{$\sqrt{k}$-approximation for Combinatorial Auction with Single-minded Bidders}
\label{sqrt_k_approximation}
\begin{algorithmic}[1]
\REQUIRE $\theta=\{(v_i,S_i^\ast)\}_{i\in N}$, $W=\emptyset$;\\
\ENSURE Allocation $f$;\\
\STATE Reorder all bids in $N$ by $ \frac{v_1}{\sqrt{|S_1^\ast|}}\geq \frac{v_2}{\sqrt{|S_2^\ast|}} \geq \cdots \frac{v_n}{\sqrt{|S_n^\ast|}}$;\\
\FOR{$i$ from $1$ to $n$}
\IF{$S_i^\ast \cap (\bigcup_{j\in W}S^\ast_j)=\emptyset$}
\STATE Update $W\leftarrow W\cup \{i\}$;\\
\ENDIF
\ENDFOR
\STATE Return the winner set $W$.
\end{algorithmic}
\end{algorithm}   

\begin{algorithm}[!htbp]
\caption{Allocation Rule of NSA Mechanism}
\label{mechanism_for_single_minded_bidders}
\begin{algorithmic}[1]
\REQUIRE $G=(N\cup\{s\}, E)$, $\theta$, $\mathcal{K}$;
\ENSURE Allocation $f$;
\STATE Initialize winner set $W$;
\STATE $\mathcal{O}\leftarrow \texttt{BFS}(G,s)$; Create the IDT $T$;
\FOR{Bidder $i$ in $\mathcal{O}$}
\STATE $N_{-T_i}\leftarrow \left(N \setminus T_i\right) \cup \{i\}$;
\IF{$i=\underset{j \in N_{-T_i}}{\arg\max} \left\{ \frac{v_j}{\sqrt{|S_j^*|}} \right\}$ and $S_i^\ast \cap (\bigcup_{j\in W}S^\ast_j)=\emptyset$}
\STATE Update $W\leftarrow W \cup\{i\}$;
\ENDIF
\ENDFOR
\STATE Return $f$ which gives $S_i^\ast$ to $i$ if and only if $i\in W$.
\end{algorithmic}
\end{algorithm}

According to \Cref{remark_2}, a straightforward way to finding a monotone allocation is to apply \Cref{sqrt_k_approximation} to all the bidders in \( N \), in conjunction with the revenue-maximizing payment scheme described in \Cref{opt_payment_id_monotonicity}. We term this mechanism  ``Network-\(\sqrt{k}\)-Approximation Mechanism (Net-\(\sqrt{k}\)-APM)''.

\begin{theorem}\label{sqrt_k_approx_mechanism}
    Net-\(\sqrt{k}\)-APM is $\sqrt{k}$-EF, IR, SP, but not WBB. 
\end{theorem}

We also consider another mechanism, termed Network Single-minded Auction (NSA) mechanism, which combines a non-trivial extension of \Cref{sqrt_k_approximation}, presented in \Cref{mechanism_for_single_minded_bidders}, along with the payment scheme introduced in \Cref{opt_payment_IP_monotonicity}.

\begin{theorem}\label{single_minded_IP_MON}
     NSA mechanism satisfies IR, SP, and WBB.
\end{theorem}

\section{Discussion}
With the characterization in the above sections, the combinatorial network auction with single-minded bidders (including the multi-unit network auctions with single-unit demand), which has been a major obstacle in the field of network auctions since 2018, is now solved in principle. Building on these insights, this work pioneers the investigation into combinatorial network auctions with single-minded bidders.
A significant open question is whether, given any value-monotone allocation rule, there always exists a computationally tractable payment rule that ensures the mechanism is strategyproof. Other intriguing questions include characterizing Bayesian truthful mechanisms, extending the deterministic 0-1 allocation in a more general context, and more.

\section*{Acknowledgments}
The authors are sincerely grateful to the reviewers for their valuable comments throughout every stage of this paper.
This research is supported by the National Natural Science Foundation of China under  Nos.  71601029, 62202229, 62372095, 62172077, and 62350710215. Dong Hao, Mingyu Xiao, and Bakh Khoussainov are supported by the Sichuan Science and Technology Program (2025HJPJ0006). Yuhang Guo is supported by the TFS Scholarship (RSRE7059 and RSRE7092) at UNSW Sydney. This work was completed during Yuhang Guo’s master’s studies at the University of Electronic Science and Technology of China.

\bibliographystyle{named}
\bibliography{ijcai25}

\cleardoublepage

\appendix

\section{Omitted Related Work} \label{sec::related_work}
\citet{LHZ+17a} initiates the study of network auctions. They showed that the classic VCG mechanism \citep{VICK61a,CLAR71a,GROV73a} can be naturally extended to networks. However, this extension can lead to low or even negative revenues. To overcome this deficit, the Information Diffusion Mechanism (IDM) was introduced. The following years saw the emergence of several network auction mechanisms with different objectives \citep{LHZ+19a,LHG+22a}. Recent progress in network auctions is reviewed in \citep{GUHA21a,ZHAO21a}. Most of these mechanisms focus on single-item network auctions. 
Some works focus on network auctions for weighted graphs \citep{LHZ+19a,LHZ24a}. \citet{XSK22a} and \citet{LWL+21a} study network auctions with budgets; double auctions with social interactions \citep{LCZ24a} have also been studied recently. \citet{FZL+24a} generalizes the design idea from \citep{FZL+23a} to a new class of mechanisms termed \emph{MetaMSN} mechanisms and explore different network combinatorial auction scenarios. \citet{SHHA22a} study all-pay auctions for contest design in social networks. Other work also studies non-truthful mechanisms for network auctions \citep{SEJO24a}.
Most of this existing work is scenario or task oriented and lacks a general theory to characterize strategyproofness. 

\section{Omitted Contents from Section 3}

\subsection{Running procedures of Table 1}
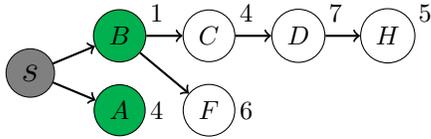
\begin{figure}[!htbp]
    \centering
\scalebox{1}{
\begin{tikzpicture}[global scale=0.99, box/.style={circle, draw}]
            \node[box,fill=gray,very thin](s) at(0,-0.5){{\LARGE$s$}};
            \node[box,fill={rgb:red,1;green,180;blue,80}](A) at(1.2,-1){$A$};
            \node[] at (1.7, -1){$4$};
            \node[box,fill=white](F) at(2.4,-1){$F$};
            \node[] at (2.9, -1){$6$};
            \node[box,fill={rgb:red,1;green,180;blue,80}](B) at(1.2,0){$B$};
            \node[] at (1.7, 0.3){$1$};
            \node[box,fill=white](C) at(2.4,0){$C$};
            \node[] at (2.9, 0.3){$4$};
            \node[box,fill=white](D) at(3.6,0){$D$};
            \node[] at (4.1, 0.3){$7$};
            \node[box,fill=white](H) at(4.8,0){$H$};
            \node[] at (5.3, 0.3){$5$};
            \draw[->,  line width=.8pt] (s) --(A);
            \draw[->,  line width=.8pt]  (s) --(B);
            \draw[->,  line width=.8pt]  (B) --(C);
            \draw[->,  line width=.8pt]  (C) --(D);
            \draw[->,  line width=.8pt]  (D) --(H);
            \draw[->,  line width=.8pt]  (B) --(F);
        \end{tikzpicture}}
\caption{Network auction market with profile $\theta$, the seller $s$ has $3$ unit items for sale.}
\label{counterexample_1_dna_mu}
\end{figure}

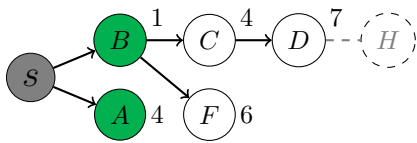
\begin{figure}[!htbp]
    \centering
\scalebox{1}{
\begin{tikzpicture}[global scale=0.99, box/.style={circle, draw}]
            \node[box,fill=gray,very thin](s) at(0,-0.5){{\LARGE$s$}};
            \node[box,fill={rgb:red,1;green,180;blue,80}](A) at(1.2,-1){$A$};
            \node[] at (1.7, -1){$4$};
            \node[box,fill=white](H) at(2.4,-1){$F$};
            \node[] at (2.9, -1){$6$};
            
            \node[box,fill={rgb:red,1;green,180;blue,80}](B) at(1.2,0){$B$};
            \node[] at (1.7, 0.3){$1$};
            \node[box,fill=white](C) at(2.4,0){$C$};
            \node[] at (2.9, 0.3){$4$};
            \node[box,fill=white](D) at(3.6,0){$D$};
            \node[] at (4.1, 0.3){$7$};
            \node[box,fill=white, dashed](H) at(4.8,0){\color{gray}{$H$}};      
            \draw[->,  line width=.8pt] (s) --(A);
            
            \draw[->,  line width=.8pt]  (s) --(B);
            \draw[->,  line width=.8pt]  (B) --(C);
            \draw[->,  line width=.8pt]  (C) --(D);
            \draw[-, dashed, gray, line width=.8pt]  (D) --(H);
            \draw[->,  line width=.8pt]  (B) --(F);
        \end{tikzpicture}}

\caption{Bidder $D$ misreport $r_D^\prime=\emptyset$, blocking bidder $H$ to enter the market}
\label{counterexample_2_dna_mu}
\end{figure}

We first provide the detailed running procedures for the counterexample with $|\mathcal{K}|=3$. Consider two different invitation strategies by bidder $D$ shown in \Cref{counterexample_1_dna_mu} and \Cref{counterexample_2_dna_mu}.

We first decide the priority order $\mathcal{O}=(A,B,F,C,D,H)$ by \texttt{BFS}. Since the markets $G$ in \Cref{counterexample_1_dna_mu} and \Cref{counterexample_2_dna_mu} are tree-structured markets, we can get the IDT directly: $T=G$. Starting from bidder $A$, since $v^3(N\setminus (T_A\cup W))=5 > v_A$, $A$ is ineligible to be selected into $W$. For bidder $B$, $v^3(N\setminus(T_B\cup W))=0 < v_B$. $B$ is qualified to be a winner and added into $W$ and pays $0$. Upon $B$ is selected, DNA-MU updates $W$ and $k$. Similarly, we can check the winning condition for the remaining bidders and the case when $D$ misreports the invitation. Details are provided in \Cref{dna_mu_table1_appendix} (for \Cref{counterexample_1_dna_mu}) and \Cref{dna_mu_table2_appendix} (\Cref{counterexample_2_dna_mu}), respectively. 

\begin{table}[!htbp]
\centering
\caption{DNA-MU Mechanism in \Cref{counterexample_1_dna_mu}}
\begin{tabular}{K{1cm}K{2cm}K{3cm}}
\toprule
   $k$  &  $W$ & $(f,p)$  \\
\midrule
   $3$  &  $\emptyset$ & $f_A=0, p_A=0$ \\
\midrule 
   $3$  &  $\emptyset$ & $f_B=1, p_B=0$ \\
\midrule 
   $2$  &  $\{B\}$     & $f_F=1, p_F=5$ \\
\midrule 
   $1$  &  $\{B,F\}$   & $f_C=1, p_C=4$ \\
\midrule 
   $0$  &  $\{B,F,C\}$ & Finished \\
\bottomrule
\end{tabular}
\label{dna_mu_table1_appendix}   
\end{table}

\begin{table}[!htbp]
\centering
\caption{DNA-MU Mechanism in \Cref{counterexample_2_dna_mu}}
\begin{tabular}{K{1cm}K{2cm}K{3cm}}
\toprule
   $k$  &  $W$ & $(f,p)$  \\
\midrule
   $3$  &  $\emptyset$ & $f_A=1, p_A=4$ \\
\midrule
   $2$  &  $\{A\}$     & $f_B=1, p_B=0$ \\
\midrule
   $1$  &  $\{A,B\}$     & $f_F=0, p_F=0$ \\
\midrule
   $1$  &  $\{A,B\}$     & $f_C=0, p_C=0$ \\
\midrule 
   $1$  &  $\{A,B\}$   & $f_D=1, p_D=6$ \\
\midrule
   $0$  &  $\{A,B,D\}$ & Finished \\
\bottomrule
\end{tabular}
\label{dna_mu_table2_appendix}   
\end{table}

\subsection{Proof of \Cref{theorem_dna_mu_fail_ic}}
\begin{proof}
From the counterexample in \Cref{counterexample_1_dna_mu} and \Cref{counterexample_2_dna_mu}, it is not hard to verify that bidder $D$ can manipulate the invitation strategy by not inviting $H$ and then become a winner, gaining extra benefits, i.e., the utility $u_D(\{H\})=0$ while $u_D(\emptyset)=1$. When $|\mathcal{K}|> 3$, we construct the counterexample in \Cref{counterexample_construction} based on \Cref{counterexample_1_dna_mu}.  
\begin{figure}[!htbp]
    \centering
    \scalebox{1}{
\begin{tikzpicture}[global scale=0.99, box/.style={circle, draw}]
            \node[box,fill=gray,very thin](s) at(0,-0.5){{\LARGE$s$}};
            \node[box,fill={rgb:red,1;green,180;blue,80}](A) at(1.2,-1){$A$};
            \node[] at (1.7, -1){$4$};
            \node[box,fill=white](F) at(2.4,-1){$F$};
            \node[] at (2.9, -1){$6$};
            
            \node[box,fill={rgb:red,1;green,180;blue,80}](B) at(1.2,0){$B$};
            \node[] at (1.7, 0.3){$1$};
            \node[box,fill=white](C) at(2.4,0){$C$};
            \node[] at (2.9, 0.3){$4$};
            \node[box,fill=white](D) at(3.6,0){$D$};
            \node[] at (4.1, 0.3){$7$};
            \node[box,fill=white](H) at(4.8,0){$H$};
            \node[] at (5.3, 0.3){$5$};

            \node[box,fill={rgb:red,1;green,180;blue,80}](P) at(-1.2,0.5){$m_1$};
            \node(100) at(-2.0,0.5) {$\bar{v}$};
            
            \node(V) at(-1.2,-.5) {\large$\vdots$};
            
            \node[box,fill={rgb:red,1;green,180;blue,80}](Q) at(-1.2,-1.5){$m_{t}$};
            \node(100) at(-2.0,-1.5) {$\bar{v}$};

            \node[rotate = 270] at (-2.6,-.5) {$\underbrace{\hspace{2.5cm}}_{|\mathcal{K}|-3 \text{ bidders}}$};
            \draw[->,  line width=.8pt] (s) --(A);
            
            \draw[->,  line width=.8pt]  (s) --(B);
            \draw[->,  line width=.8pt]  (B) --(C);
            \draw[->,  line width=.8pt]  (C) --(D);
            \draw[->,  line width=.8pt]  (D) --(H);
            \draw[->,  line width=.8pt]  (B) --(F);
            \draw[->,  line width=.8pt]  (s) --(P);
            \draw[->,  line width=.8pt]  (s) --(V);
            \draw[->,  line width=.8pt]  (s) --(Q);     
        \end{tikzpicture}}
    \caption{Construction of counterexample where $|\mathcal{K}|>3$.}
    \label{counterexample_construction}
\end{figure}
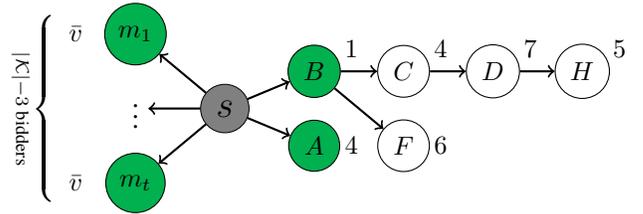

The construction keeps the original network auction market with $\{A,B,C,F,D,H\}$ and creates $t=|\mathcal{K}|-3$ dummy bidders (nodes $m_1,\cdots, m_t$ in \Cref{counterexample_construction}), directly connecting to the seller $s$ with bid $\bar{v} > \max\{v_A,v_B,v_C, v_F, v_D,v_H\}$. By the DNA-MU mechanism, these $t$ bidders are selected as the first $t$ winners. After the allocation of the first $|\mathcal{K}|-3$ units, it degenerates to the original counterexample that we construct where $|\mathcal{K}|=3$. This completes the proof that DNA-MU mechanism fails to satisfy SP when $|\mathcal{K}|\geq 3$.
\end{proof}

\subsection{DNA-MU-R Mechanism}
We first demonstrate the DNA-MU-R mechanism in an standard algorithm form in \Cref{FIX_DNA_MU} and revisit the counterexample in \Cref{counterexample_1_dna_mu} and \Cref{counterexample_2_dna_mu}.

\begin{algorithm}[!htbp]
\caption{DNA-MU-Refined (DNA-MU-R) Mechanism}
\label{FIX_DNA_MU}
\begin{algorithmic}[1]
\REQUIRE  $G=(N\cup\{s\}, E)$, $\mathbf{\theta}$, $\mathcal{K}$;\\
\ENSURE  Allocation $f$, payment $p$;
\STATE Initialize order $\mathcal{O}\leftarrow \mathtt{BFS}(G, s)$;\\
\STATE Create Invitational-Domination Tree (IDT) $T$;\\
\STATE Initialize $k \leftarrow |\mathcal{K}|, W\leftarrow \emptyset$;\\
\FOR{$i$ in $\mathcal{O}$}
\STATE $T_i\leftarrow $ Sub-Tree rooted by $i$ in $T$;\\
\IF{{\color{red}{$v_i \geq v^{k}(N\setminus T_i)$}}}
\STATE $f_i\leftarrow 1,{\color{red}{p_i\leftarrow v_i^\ast(r_i)}} $;\\
\STATE Update $ W\leftarrow W\cup \{i\}$;\\
\ENDIF
\ENDFOR
\STATE \textbf{Return} $f, p$.
\end{algorithmic}
\end{algorithm}

The detailed running procedures of DNA-MU-R mechanism in \Cref{counterexample_1_dna_mu} and \Cref{counterexample_2_dna_mu} is illustrated in \Cref{dna_mu_r_table1_appendix} and \Cref{dna_mu_r_table2_appendix}. It is not hard to see that bidder $D$ has no incentive to deviate as no matter what invitation strategy is taken by $D$, $D$ is unable to gain extra benefits. 

\begin{table}[!htbp]
\centering
\caption{DNA-MU-R Mechanism in \Cref{counterexample_1_dna_mu}}
\begin{tabular}{K{2cm}K{1.5cm}K{3cm}}
\toprule
   Left Units  &  $W$ & $(f,p)$  \\
\midrule
   $3$  &  $\emptyset$ & $f_A=0, p_A=0$ \\
\midrule 
   $3$  &  $\emptyset$ & $f_B=1, p_B=0$ \\
\midrule 
   $2$  &  $\{B\}$     & $f_F=1, p_F=4$ \\
\midrule 
   $1$  &  $\{B,F\}$   & $f_C=1, p_C=1$ \\
\midrule 
   $0$  &  $\{B,F,C\}$ & Finished \\
\bottomrule
\end{tabular}
\label{dna_mu_r_table1_appendix}   
\end{table}

\begin{table}[!htbp]
\centering
\caption{DNA-MU-R Mechanism in \Cref{counterexample_2_dna_mu}}
\begin{tabular}{K{2cm}K{1.5cm}K{3cm}}
\toprule
   Left Units  &  $W$ & $(f,p)$  \\
\midrule
   $3$  &  $\emptyset$ & $f_A=1, p_A=4$ \\
\midrule 
   $2$  &  $\{A\}$ & $f_B=1, p_B=0$ \\
\midrule 
   $1$  &  $\{A,B\}$     & $f_F=1, p_F=4$ \\
\midrule 
   $0$  &  $\{A,B,F\}$   & Finished \\
\bottomrule   
\end{tabular}
\label{dna_mu_r_table2_appendix}   
\end{table}

\subsection{Proof of \Cref{DNA_MU_R_properties}}
\begin{proof}
To prove IR and SP, we show all the four axioms in \Cref{ic_diffusion_auction_theorem} are satisfied by the DNA-MU-R mechanism $M=(f,p)$. 

\textbf{Value-monotone allocation}: For each bidder $i$, if $i$ is a winner, it means when $i$ is checked to determine whether she can be a winner or not, there must exist some unit unallocated and $v_i \geq v^k(N\setminus T_i)$. Misreporting high bid $v_i^\prime > v_i$ could only cause some higher priority bidders lose the auction, rising the unallocated count when $i$ being checked. Also, $v_i^\prime > v_i \geq v^k(N\setminus T_i)$. Thus, $i$ will still be a winner. Therefore, the allocation function $f$ of DNA-MU-R mechanism is value-monotone.

\textbf{Bid-independent and invitational-monotone payment}:
The payment for each bidder \( i \) is bid-independent according to the definition of \( v^\ast_i(r_i) \). For invitational-monotonicity, note that the payment rule of the DNA-MU-R mechanism can be considered as \( \tilde{p}_i(r_i) = v_i^\ast(r_i) \) and \( \bar{p}_i(r_i) = 0 \). Obviously, \( \bar{p}_i(r_i) \) is invitational-monotone. For \( \tilde{p}_i \), it suffices to prove that for any \( r_i^\prime \subseteq r_i \), \( v_i^\ast(r_i) \leq v_i^\ast(r_i^\prime) \).

According to the definition of \( v_i^\ast(r_i) \), this means proving that for any \( v_i^\prime \) such that \( f_i((v_i^\prime, r_i^\prime), \theta_{-i}) = 1 \), we always have \( f_i((v_i^\prime, r_i), \theta_{-i}) = 1 \). To facilitate the proof, we first define \( k_i \) and \( k_i^\prime \), where \( k_i \) is the number of unallocated units when checking \( i \) with the reported type \( (v_i^\prime, r_i) \) and \( k_i^\prime \) for the reported type \( (v_i^\prime, r_i^\prime) \). We first have that 

According to the DNA-MU-R mechanism, if \( i \) reports \( (v_i^\prime, r_i^\prime) \) and becomes a winner, then we must have \( k_i^\prime > 0 \) and \( v_i^\prime \geq v^{k}(N \setminus T_i) \). Now consider reporting \( (v_i^\prime, r_i) \). Since \( v_i^\prime \) is unchanged, \( v_i^\prime \geq v^k(N \setminus T_i) \) still holds. By inviting more neighbors in \( r_i \setminus r_i^\prime \), for each bidder \( j \) who has higher priority than \( i \), it only becomes harder for \( j \) to be a winner since \( v^k(N \setminus T_j) \) is non-decreasing with a larger set \( N \setminus T_j \).

The only case we need to discuss is when some previous winners with \( r_i^\prime \) become losers with \( r_i \), and some bidder \( q \) whose priority is between \( j \) and \( i \) ends up with a higher \( k_q \) and thus becomes a winner. In this case, there could be only two different situations: (1) bidder \( j \) becomes a loser with \( r_i \), and no such bidder \( q \) exists; (2) bidder \( j \) becomes a loser with \( r_i \), and only one such bidder \( q \) changes from a loser to a winner.

Assume there are two bidders \( q_1 \) and \( q_2 \) who are both losers with \( r_i^\prime \) and become winners with \( r_i \). We discuss the following four different cases to show the impossibility that \( q_1 \) and \( q_2 \) can be winners simultaneously.
\begin{itemize}
    \item Both $q_1$ and $q_2$ invitationally dominate $i$: then $v^k(N\setminus T_{q_1})$ and $v^k(N\setminus T_{q_2})$ keep the same for $r_i$ and $r_i^\prime$. W.l.o.g, assume $q_1$ is prior than $q_2$, if $q_1$ becomes a winner for $r_i$, then the unallocated unit number $k_{q_2}=k_{q_2}^\prime + 1 -1=k_{q_2}^\prime$, keeping the same. Therefore, $q_2$ will still be a loser in this case.
    \item $q_1$ invitationally dominates $i$ and $q_2$ does not dominate $i$. In this case, if $q_1$ becomes a winner, $k_{q_2}$ keeps the same with $r_i$ and $r_i^\prime$, also, since $q_2$ does not dominate $i$, $v^k(N\setminus T_{q_2})$ could become larger, making $q_2$ harder to a winner. Thus, $q_2$ still loses.
    \item $q_1$ does not invitationally dominate $i$ and $q_2$ dominates $i$. If $q_1$ becomes a winner, the winning condition for $q_2$ is not changed as illustrated in the first situation.
    \item Both $q_1$ and $q_2$ do not invitationally dominate $i$: If $q_1$ becomes a winner, then winning condition for $q_2$ becomes harder as illustrated in the second situation.
\end{itemize}

Now, we have shown the fact that ``if there exists a bidder $j$ changes from winner to loser because of $i$ misreports $r_i^\prime$ to $r_i$, then at most one bidder $q$ whose priority is between $j$ and $i$ changes from a loser to a winner." This means when it comes to $i$ with $(v_i^\prime, r_i)$, the unallocated unit number $k_i \geq k_i^\prime > 0$, $i$ keeps to be a winner with reporting type $(v_i^\prime, r_i)$. i.e., $\forall \, v_i^\prime$ such that $f_i((v_i^\prime, r_i^\prime), \theta_{-i})$, $f_i((v_i^\prime, r_i),\theta_{-i})=1$. This implies that $v^\ast_i(r_i) \leq v^\ast_i(r_i^\prime)$, i.e., $\tilde{p}_i(r_i) \leq \tilde{p}_i(r_i^\prime)$.

\textbf{Condition 3 and 4 in \Cref{ic_diffusion_auction_theorem}}: notice that $\tilde{p}_i(r_i)=v^\ast_i(r_i)$ and $\bar{p}_i(r_i)=0$, then it naturally holds for $\tilde{p}_i(r_i) - \bar{p}_i(r_i) = v^\ast_i(r_i)$ and $\bar{p}_i(\emptyset)\leq 0$.

With regard to the property of WBB, $\Rev^{\mathcal{M}}(\theta)=\sum_{i\in N}p_i(r_i)=\sum_{i: f_i=1}v^\ast_i(r_i)\geq 0$.
\end{proof}

\subsection{Justification of payment $v_i^\ast(r_i)$ in DNA-MU-R Mechanism}\label{example_dna_mu_r}

We justify that it is essential to express the payment rule in DNA-MU-R mechanism by $v^\ast_i(r_i)$, rather than an explicit form by considering the following instance.
\begin{figure}[!htbp]
    \centering
    \scalebox{1}{
\begin{tikzpicture}[global scale=0.99, box/.style={circle, draw}]
            \node[box,fill=gray,very thin](s) at(0,-0.5){{\LARGE$s$}};
            \node[box,fill={rgb:red,1;green,180;blue,80}](A) at(1.2,-1){$A$};
            \node[] at (1.7, -0.7){$3$};
            \node[box,fill=white](F) at(2.4,-1){$F$};
            \node[] at (2.9, -0.7){$2$};
            \node[box,fill={rgb:red,1;green,180;blue,80}](B) at(1.2,0){$B$};
            \node[] at (1.7, 0.3){$1$};
            \node[box,fill=white](C) at(2.4,0){$C$};
            \node[] at (2.9, 0.3){$2$};
            \node[box,fill=white](D) at(3.6,0){$D$};
            \node[] at (4.1, 0.3){$4$};
            \node[box,fill=white](H) at(4.8,0){$H$};
            \node[] at (5.3, 0.3){$6$};
            \node[box,fill=white](G) at(4.8,-1){$G$};
            \node[] at (5.3, -0.7){$5$};
            \draw[->,  line width=.8pt] (s) --(A);
            \draw[->,  line width=.8pt]  (s) --(B);
            \draw[->,  line width=.8pt]  (B) --(C);
            \draw[->,  line width=.8pt]  (C) --(D);
            \draw[->,  line width=.8pt]  (D) --(H);
            \draw[->,  line width=.8pt]  (A) --(F);
            \draw[->,  line width=.8pt]  (D) --(G);
        \end{tikzpicture}}
\caption{Multi-unit network auction with unit demand bidders. Seller $s$ has $3$ units for sale.}
\label{example_2_dna_mu_r_appendix} 
\end{figure}
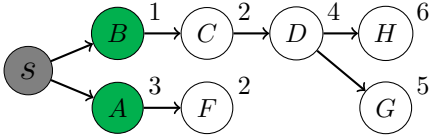

We first decide the priority order $\mathcal{O} = (A,B,F,C,D,G,H)$ and the invitational domination tree (IDT) $T=G$. For bidder $A$, $v^3(N\setminus T_A)=4 > v_A=3$. $A$ is not eligible to be a winner $f_A=0$ and pays zero $p_A=0$; For bidder $B$, since $v^3(N\setminus T_B)=0 < v_B=1$, $B$ is added into $W$ and pays the critical bid $0$. Similarly, for bidder $F$, we have that $F$ is not qualified to be a winner as $v^3(N\setminus T_F)=4 > v_F=2$ and $F$ pays $0$; For bidder $C$, $v^3(N\setminus T_C)=1 < v_C=2$ and $C$ pays her critical bid $1$. It is worth noting that for bidder $D$, we have $v^3(N\setminus T_D)=2 < v_D=4$, implying $D$ is eligible to be a winner. However, the critical bid of bidder $D$ is $3$ rather than $v^3(N\setminus T_D)=2$. Assuming that bidder $D$ reports $v_D=2$, $D$ will lose the auction since the final winners will be $\{A,B,C\}$ according to the allocation of DNA-MU-R mechanism. This example shows that critical bid $v_i^\ast(r_i)$ could be inconsistent with $v^k(N\setminus T_i)$ because of the allocation rule of DNA-MU-R mechanism. This justifies the necessity of utilizing $v^\ast_i(r_i)$ as the payment expression.

\section{Omitted Contents from Section 4}
\subsection{Proof of \Cref{ic_payment_scheme}}
\begin{proof}
Given any value-monotone allocation function $f$, to prove that the above construction of payment function $p$ makes $(f,p)$ strategyproof, we show that such a pair $(f,p)$ satisfies all the premises in \Cref{ic_diffusion_auction_theorem}.

Firstly, $f$ is value-monotone. Next, $\tilde{p}_i(r_i) - \bar{p}_i(r_i) = \tilde{g}(r_i) - \bar{g}(r_i) = v^\ast_i(r_i)$. Also, both $\tilde{p}_i$ and $\bar{p}_i$ are bid-independent because $\tilde{g}(r_i)$, $\bar{g}(r_i)$, and $h(\theta_{-i})$ are independent with $v_i$. For the invitational-monotonicity of payment, taking $\tilde{p}$ as the example, for any bidder $i$ and any invitation strategy $r_i^1\subseteq r_i^2$, we show $\tilde{p}_i(r_i^1) \geq \tilde{p}_i(r_i^2)$. 
Consider the profile $\theta_i=(v_i,r_i^2)$, according to the construction, we always have $r_i^2=\arg\min_{r_i^\prime\subseteq r_i^2} \tilde{g}(r_i^\prime) $, which implies $\tilde{g}(r_i^2) \leq \tilde{g}(r_i^1)$. For any two invitation strategies $r_i^1\subseteq r_i^2$, this payment construction always guarantees that $\tilde{g}(r_i^2) \leq \tilde{g}(r_i^1)$. This represents the invitational-monotonicity for $\tilde{p}$. Analogously, we can prove the invitational-monotonicity for $\bar{p}$.
\end{proof}

\subsection{Proof of \Cref{critial_bid_with_allocation}}
\begin{proof}
    $(\implies)$ For bidder $i$, since $v_i^\ast(r_i^1) \leq v_i^\ast(r_i^2)$, the bidding space $\mathcal{R}_{\geq 0}$ is divided into three intervals: $[0, v_i^\ast(r_i^1))$, $[v_i^\ast(r_i^1), v_i^\ast(r_i^2))$, and $[v_i^\ast(r_i^2),+\infty)$. Note that $f$ is value monotone, (1). $v_i\in [0, v_i^\ast(r_i^1))$, $f(v_i,r_i^1) = f(v_i,r_i^2)=0$; (2). $v_i\in [v_i^\ast(r_i^1), v_i^\ast(r_i^2))$, $f(v_i,r_i^1)=1,f(v_i,r_i^2)=0$; (3). $v_i \in [v_i^\ast(r_i^2),+\infty)$, $f(v_i,r_i^1)=f(v_i,r_i^2)=1$. Thus, for any $v_i\in \mathcal{R}_{\geq 0}$, $f(v_i,r_i^1) \geq f(v_i,r_i^2)$. \\
    $(\impliedby)$ For bidder $i$, if it holds that $\forall~ v_i\in \mathcal{R}_{\geq 0}$, $f(v_i,r_i^1) \geq f(v_i,r_i^2)$,
    then $f(v^\ast(r_i^1),r_i^1) \geq f(v^\ast(r_i^1),r_i^2)$ and $f(v^\ast(r_i^2),r_i^1) \geq f(v^\ast(r_i^2),r_i^2)$, according to the definition of $v^\ast(\cdot)$, $f(v^\ast(r_i^1),r_i^1)=1$ and $f(v^\ast(r_i^2),r_i^2)=1$, Note that $f(v^\ast(r_i^1),r_i^2) \leq 1$ and $f(v^\ast(r_i^2),r_i^2) = 1$, then $v^\ast(r_i^1) \leq v^\ast(r_i^2)$ since $f$ is value-monotone.
\end{proof}

\subsection{Proof of \Cref{critical_bid_monotonicity_competitive}}
\begin{proof}
    According to the definition of ID-MON partial ordering, $\theta_i^1 \succeq_{\mathcal{D}} \theta_i^2$, which means $f(\theta_i^1, \theta_{-i}) \geq f(\theta_i^2,\theta_{-i})$. Note that for any fixed bid $v_i$, we always have $\theta_i^1 \succeq_{\mathcal{D}} \theta_i^2$, that is to say, $\forall \, v_i \in \mathcal{R}_{\geq 0}, f(v_i,r_i^1) \geq f(v_i,r_i^2)$. According to \Cref{critial_bid_with_allocation}, this implies $v_i^\ast(r_i^1) \leq v_i^\ast(r_i^2)$.
\end{proof}

\subsection{Proof of \Cref{ic_implementability_id_mon}}
\begin{proof}
According to \Cref{ic_payment_scheme}, since ID-MON implies that $\forall \, i\in N$, $r_i^1 \subseteq r_i^2\subseteq r_i$, $v_i^\ast(r_i^1) \leq v_i^\ast(r_i^2)$, we can choose some polynomial functions for $\tilde{g}(\cdot)$ and $\bar{g}(\cdot)$, which are non-increasing with $v_i^\ast(r_i)$. For instance, let $\tilde{g}(r_i)=-\alpha v_i^\ast(r_i) - \beta (v^\ast_i(r_i))^2$, $h(\theta_{-i})=\gamma v^\ast_i(\emptyset)$ and $\bar{g}_i(r_i)=\gamma v^\ast_i(\emptyset) - (1 + \alpha) v^\ast_i(r_i) - \beta (v^\ast_i(r_i))^2$, where $\alpha,\beta,\gamma \geq 0$, then all the conditions in \Cref{ic_payment_scheme} can be satisfied. Thus, every ID-MON allocation $f$ is network-implementable.
\end{proof}

\subsection{Proof of \Cref{opt_payment_id_monotonicity}}
\begin{proof}
This proof follows the idea from Theorem 4 in \cite{LHZ20a}. Given an ID-MON allocation $f$ and bidder type profile $\theta$. The seller $s$'s revenue can be represented by
    \begin{equation}
            \Rev^{\mathcal{M}}(\theta) = \sum_{i\in N} p_i(r_i) =\sum_{i:f_i=1}\tilde{p}_i(r_i)+\sum_{j:f_j=0}\bar{p}_j(r_j).    
    \end{equation}
    In order to guarantee the mechanism $\mathcal{M}=(f,p)$ be IR and SP. The conditions (1) to (4) in \Cref{ic_diffusion_auction_theorem} should be satisfied. According to  condition (3) which is $\forall\, i\in N, \tilde{p}_i(r_i)-\bar{p}_i(r_i)=v^\ast_i(r_i)$, we can rewrite the revenue as:
    \begin{equation}
        \Rev^{\mathcal{M}}(\theta)=\sum_{i:f_i=1}v^\ast_i(r_i)+\sum_{j\in N} \bar{p}_j(r_j).
    \end{equation}
    Note that when $f$ is given, for each bidder $i$, the critical winning bid $v^\ast_i(r_i)$ is a constant value for each given profile $\theta$. Thus, in order to maximize the revenue, we should optimize the payment rule $p$ to maximize $\sum_{i\in N}\bar{p}_i(r_i)$. Furthermore, it is not hard to see that bidders' losing payment $\bar{p}_i$ are pairwisely independent (This means when the payment rule is decided, for any bidders $i$ and $j$, $\tilde{p}_i$ (resp. ${\bar{p}}_i$) is determined by the profile, which is independent with $\tilde{p}_j$ (resp. ${\bar{p}}_j$)). Then the problem of maximizing the global revenue can be transformed into maximizing each bidder's losing payment value:
\begin{subequations}
    \begin{align}
        \max_{\bar{p}}\quad & \bar{p}_i(r_i) \nonumber \\
        \text{s.t.}\quad & \forall\, v_i^1,v_i^2 \in \mathcal{R}_{\geq 0}, v_i^1\neq v_i^2,\bar{p}_i(v_i^1,r_i) = \bar{p}_i(v_i^2, r_i)\label{3a} \\ 
        &\forall\, v_i^1,v_i^2 \in \mathcal{R}_{\geq 0}, v_i^1\neq v_i^2,\tilde{p}_i(v_i^1,r_i)=\tilde{p}_i(v_i^2,r_i) \\
        & \forall \, r_i^\prime \subseteq r_i,\tilde{p}_i(r_i^\prime) =\bar{p}_i(r_i^\prime) + v^\ast_i(r_i^\prime) \label{3b}\\
        & \forall \, r_i^\prime \subseteq r_i, \tilde{p}_i(r_i) \leq \tilde{p}_i(r_i^\prime),\bar{p}_i(r_i) \leq \bar{p}_i(r_i^\prime) \label{3c} \\
        & \bar{p}_i(\emptyset) \leq 0 \label{3d}\\
        & \forall \, r_i^\prime \subseteq r_i, v^\ast_i(r_i) \geq v^\ast_i(r_i^\prime).
        \label{3e}
    \end{align}
\end{subequations}
Constraints (\ref{3a}) to (\ref{3d}) on the optimization problem correspond to the conditions (1) to (4) in the \Cref{ic_diffusion_auction_theorem} while constraint (\ref{3e}) represents the deterministic ID allocation rule via \Cref{critical_bid_monotonicity_competitive}.  
To solve this optimization problem, we make the following deduction. According to conditions (\ref{3b}), (\ref{3c}), and (\ref{3e}), in order to satisfy the edge-monotonicity of payment, we obtain the inequality $\bar{p}_i(r_i) + v^\ast_i(r_i) \leq \bar{p}_i(r_i^\prime) + v^\ast_i(r_i^\prime),$ where $r_i^\prime \subseteq r_i$. Taking $r_i^\prime = \emptyset$ and rewriting the inequality, we get $\bar{p}_i(r_i) \leq \bar{p}_i(\emptyset) + v^\ast_i(\emptyset) - v^\ast_i(r_i).$ Condition (\ref{3d}) tells us that $\bar{p}_i(\emptyset) \leq 0$, so we obtain the upper bound on $\bar{p}_i(r_i)$, which is $v_i^\ast(\emptyset) - v^\ast_i(r_i)$. Next, by condition (\ref{3b}), we have $\tilde{p}_i(r_i) = v^\ast_i(\emptyset)$. Thus, the payment rule $\tilde{p}_i(r_i)=v^\ast_i(\emptyset)$ and $\bar{p}_i(r_i)=v_i^\ast(\emptyset) - v_i^\ast(r_i)$ maximizes the seller's revenue for any given ID-MON allocation rule with IR and SP constraints.
\end{proof}

\subsection{Proof of \Cref{VCG_RM_properties}}
\begin{proof}
The VCG-RM mechanism is efficient because it allocates the $|\mathcal{K}|$ items to bidders with the top-$|\mathcal{K}|$ bidders. To prove IR, SP and $\Rev^{\text{VCG-RM}}(\theta) \geq \Rev^{\text{VCG}}(\theta)$, we firstly prove that efficient allocation is always ID-MON and then prove the payment rule devised in VCG-RM mechanism is equal to the revenue-maximizing payment scheme in \Cref{opt_payment_id_monotonicity}.

\textbf{Efficient Allocation satisfies ID-MON}: for each bidder $i$, consider two type profiles $\theta_i^1=(v_i^1,r_i^1)$ and $\theta_2=(v_i^2,r_i^2)$, where $v_i^1 \geq v_i^2$ and $r_i^1 \subseteq r_i^2$, if $i$ reports type $(v_i^2,r_i^2)$ and becomes a winner in efficient allocation, i.e., $f_i((v_i^2,r_i^2), \theta_{-i})=1$, then it means $v_i^2$ is in the top-$k$ highest bids among all the bidders in $N\setminus r_i\cup r_i^2$. Since $v_i^1\geq v_i^2$ and $(N\setminus r_i\cup r_i^1) \subseteq (N\setminus r_i\cup r_i^2)$, $v_i^1$ is still in the top-$k$ highest bids among bidders in $N\setminus r_i\cup r_i^1$, thus if $i$ reports type $(v_i^1,r_i^1)$, $f_i((v_i^1,r_i^1), \theta_{-i})=1$. This implies the efficient allocation rule satisfies ID-MON.

\textbf{Revenue-Maximizing Payment Rule}: For $\tilde{p}_i(r_i)=v^k(N\setminus T_i)$, winner $i$ pays $k$-th highest bid among $N\setminus T_i$. According to the definition of $v^\ast_i(\emptyset)$, when $i$ reports $r_i^\prime =\emptyset$, if $i$ is a winner, then her bid $v_i$ is in the top-$k$ among $N\setminus \{T_i\} \cup \{i\}$, the minimum value that $i$ can bid to maintain her winner position is the $k$-th highest bid among $N\setminus T_i$. That is $v_i^\ast(\emptyset)=v^k(N\setminus T_i)$. The same deduction we can see $v_i^\ast(r_i) = v^k(N)$. Then for VCG-RM mechanism, $\tilde{p}_i(r_i)=v^k(N\setminus T_i) = v^\ast_i(\emptyset)$ and $\bar{p}_i(r_i)=v^k(N)$. 
\end{proof}
\subsection{Proof of \Cref{critial_bid_IP_monotonicity}}
\begin{proof}
    According to the definition of IP partial ordering, $\theta_i^1 \succeq_{\mathcal{P}} \theta_i^2$, which means $f(\theta_i^1, \theta_{-i}) \geq f(\theta_i^2,\theta_{-i})$. Note that for any fixed bid $v_i$, we always have $\theta_i^1 \succeq_{\mathcal{P}} \theta_i^2$, that is to say, $\forall \, v_i \in \mathcal{R}_{\geq 0}, f(v_i,r_i^1) \geq f(v_i,r_i^2)$. According to \Cref{critial_bid_with_allocation}, this implies $v_i^\ast(r_i^1) \leq v_i^\ast(r_i^2)$.
\end{proof}

\subsection{Proof of \Cref{ic_implementability_ip_mon}}

\begin{proof}
    The proof idea is similar to that of \Cref{ic_implementability_id_mon}, according to \Cref{ic_payment_scheme}, since IP-MON implies that $\forall \, i\in N$, $r_i^2 \subseteq r_i^1\subseteq r_i$, $v_i^\ast(r_i^1) \leq v_i^\ast(r_i^2)$, we can choose some polynomial functions for $\tilde{g}(\cdot)$ and $\bar{g}(\cdot)$ and the function is non-decreasing with $v_i^\ast(r_i)$. For instance, let $\tilde{g}(r_i)=\alpha v_i^\ast(r_i) + \beta (v^\ast_i(r_i))^2$. By choosing $h(\theta_{-i})=\gamma v^\ast_i(\emptyset)$ and $\bar{g}(r_i)=\gamma v^\ast_i(\emptyset) + (\alpha - 1) v^\ast_i(r_i) + \beta (v^\ast_i(r_i))^2$, where $\alpha,\beta,\gamma \geq 0$, then all the conditions described in \Cref{ic_payment_scheme} can be satisfied. Thus, any IP-MON allocation rule $f$ is network-implementable.
\end{proof}

\subsection{Proof of \Cref{opt_payment_IP_monotonicity}}
\begin{proof}
    Notice that neither ID-MON nor IP-MON allocation rule influences the reduction process from global revenue maximization problem to $n$ independent sub-optimization problems, so the maximization goal is still $\bar{p}_i(r_i)$. Moreover, the only difference in building up the optimization problem for maximize $\bar{p}_i(r_i)$ is regarding the IP-MON, more specifically, the critical winning bid monotonicity: $\forall\, r_i^\prime \subseteq r_i, v_i^\ast(r_i)\geq v_i^\ast(r_i^\prime)$. Formally, 
\begin{subequations}
    \begin{align}
        \max_{\bar{p}}\quad & \bar{p}_i(r_i) \nonumber \\
        \text{s.t.}\quad & \forall\, v_i^1,v_i^2 \in \mathcal{R}_{\geq 0}, v_i^1\neq v_i^2,\bar{p}_i(v_i^1,r_i) = \bar{p}_i(v_i^2, r_i)\label{4a}\\
        & \forall\, v_i^1,v_i^2 \in \mathcal{R}_{\geq 0}, v_i^1\neq v_i^2, \tilde{p}_i(v_i^1,r_i)=\tilde{p}_i(v_i^2,r_i) \\
        & \forall \, r_i^\prime \subseteq r_i,\tilde{p}_i(r_i^\prime) =\bar{p}_i(r_i^\prime) + v^\ast_i(r_i^\prime) \label{4b}\\
        & \forall \, r_i^\prime \subseteq r_i, \tilde{p}_i(r_i) \leq \tilde{p}_i(r_i^\prime),\bar{p}_i(r_i) \leq \bar{p}_i(r_i^\prime) \label{4c} \\
        & \bar{p}_i(\emptyset) \leq 0 \label{4d}\\
        & \forall \, r_i^\prime \subseteq r_i, v^\ast_i(r_i) \leq v^\ast_i(r_i^\prime). \label{4e}
    \end{align}
\end{subequations}

    Condition (\ref{4a}) to (\ref{4d}) are the IR and SP restrictions and condition (\ref{4e}) is from the IP-Monotonicity. Since our target is to maximize $\bar{p}_i(r_i)$ and the payment monotonicity requires $\bar{p}_i(r_i) \leq \bar{p}_{i}(r_i^\prime) \leq \cdots  \leq \bar{p}_i(\emptyset) \leq 0$. Thus, $\bar{p}_i(r_i)$ is upper bounded by $0$. Next, we show that the solution $\bar{p}_i(r_i) = 0$ is feasible.  Firstly, condition (\ref{4a}) is naturally satisfied; To achieve condition (\ref{4b}), we just set $\tilde{p}_i(r_i)=v^\ast_i(r_i)$; For the winning payment monotonicity, upon we set up the $\bar{p}_i(r_i)=0$, then $\bar{p}_i(r_i) = \bar{p}_{i}(r_i^\prime) = \cdots = \bar{p}_i(\emptyset) = 0$, thus the winning payment always equals to critical winning bid. The constrain (\ref{4c}) regarding the winning payment invitational-monotonicity ($\forall\, r_i^\prime \subseteq r_i, \tilde{p}_i(r_i^\prime) \geq \tilde{p}_i(r_i)$) is directly satisfied due to the monotonicity consistence regarding critical winning bid: $\forall \, r_i^\prime \subseteq r_i, v_i^\ast(r_i^\prime) \geq v_i^\ast(r_i)$. Therefore, once an IP-MON allocation rule is given, the payment rule $\tilde{p}_i(r_i)=v_i^\ast(r_i)$ and $\bar{p}_i(r_i)=0$ maximize the seller's revenue under IR and SP constraints.
\end{proof}

\subsection{Proof of \Cref{DNA_MU_R_allocation_IP_MON}}
\begin{proof}
To prove the DNA-MU-R mechanism's allocation rule $f$ satisfies IP-MON, we show that for any bidder $i$ with reporting type $(v_i^1,r_i^1)$, if $f_i((v_i^1,r_i^1), \theta_{-i})=1$, then for any other reporting type $(v_i^2,r_i^2)$ where $v_i^2 \geq v_i^1$ and $r_i^1 \subseteq r_i^2$, $f_i((v_i^2,r_i^2),\theta_{-i})=1$. 

Since reporting $(v_i^1,r_i^1)$ makes $i$ become a winner, then when $i$ is checked, there must exists at least one unit unallocated and $v_i^1 \geq v^k(N\setminus T_i)$. Therefore, $v_i^2 \geq v_i^1 \geq v^k(N\setminus T_i)$. Now we prove when reporting $(v_i^2, r_i^2)$, there still exists at least one unit unallocated when $i$ is checked. The idea follows the proof of \Cref{DNA_MU_R_properties}, increasing to $v_i^2$ could only make some other bidders who have higher priority become losers while inviting more bidders in $r_i^2\setminus r_i^1$ could only increase the remaining unallocated unit number. Hence, we will have $f_i((v_i^2,r_i^2), \theta_{-i})=1$. This shows that DNA-MU-R mechanism's allocation rule satisfies IP-MON.
\end{proof}

\subsection{Proof of \Cref{degenerate_lemma}}
\begin{proof}
    Consider the allocation rule $f$ which is both ID-MON and IP-MON, then for each bidder $i$ and two different type profile $(v_i, r_i^1)$ and $(v_i,r_i^2)$ where $r_i^1 \subseteq r_i^2$, ID-MON implies $f_i((v_i,r_i^1), \theta_{-i}) \geq f_i((v_i,r_i^2), \theta_{-i})$ while IP-MON implies $f_i((v_i,r_i^1), \theta_{-i}) \leq  f_i((v_i,r_i^2), \theta_{-i})$. The only condition under which the two inequalities mentioned above hold is equality, i.e., $f_i((v_i,r_i^1), \theta_{-i}) = f_i((v_i,r_i^2), \theta_{-i})$. Note this holds for any invitational strategies with subset relation. Then for each bidder $i$, the allocation is only determined by her bid, independent with invitational strategy $r_i$, i.e., $\forall r_i^\prime \subseteq r_i$, $f_i((v_i,r_i^\prime),\theta_{-i})=f_i((v_i,r_i),\theta_{-i})$. This implies that for any invitational strategies $r_i^\prime \subseteq r_i$, $v^\ast_i(r_i^\prime)=v^\ast_i(r_i)$.
    Notice that if $p$ is under the scheme in \Cref{opt_payment_id_monotonicity} and \Cref{opt_payment_IP_monotonicity}, then the revenue-maximizing payment schemes are the same: for each bidder $i\in N$, we have  $\tilde{p}_i(r_i)=v_i^\ast(\emptyset)=v^\ast_i(r_i)$ and $\bar{p}_i(r_i)=v_i^\ast(\emptyset) - v_i^\ast(r_i)=0$. Now consider in any profile $\theta$, for any bidder $i$, consider bidder $i$'s two type profile: $\theta_i=(v_i,r_i)$ and $\theta_i^\prime=(v_i,r_i^\prime)$ where $r_i^\prime \subseteq r_i$, the utility in mechanism $\mathcal{M}$ for these two type profiles can be represented by:
    \[
    \begin{aligned}
         u_i(\theta_i,\theta_{-i})=&f_i((v_i,r_i),\theta_{-i})v_i - [f_i((v_i,r_i),\theta_{-i})\tilde{p}_i(r_i) \\
        & - (1-f_i((v_i,r_i),\theta_{-i}) )\bar{p}_i(r_i)]; \\
        u_i(\theta_i^\prime,\theta_{-i})=&f_i((v_i,r_i^\prime),\theta_{-i})v_i
        - [f_i((v_i,r_i^\prime),\theta_{-i})\tilde{p}_i(r_i^\prime)\\ & - (1-f_i((v_i,r_i^\prime),\theta_{-i}) )\bar{p}_i(r_i^\prime)].
    \end{aligned}
    \]
    Since $f_i((v_i,r_i),\theta_{-i}) = f_i((v_i,r_i^\prime),\theta_{-i})$ and  $v_i^\ast(r_i)=v_i^\ast(r_i^\prime)$, then we abbreviate them as $f_i$ and $v_i^\ast$, respectively. With $\tilde{p}_i=v_i^\ast,\bar{p}_i=0$, substituting the payment rule, we can obtain:
    \[
    u_i(\theta_i,\theta_{-i}) = f_i (v_i - v_i^\ast) = u_i(\theta_i^\prime, \theta_{-i}).
    \]
    Therefore, the mechanism $\mathcal{M}$ is degenerated.
\end{proof}

\subsection{Proof of \Cref{special_cases_of_lemma_2}}
\begin{proof}
For ID-MON allocations, set $\tilde{g}(r_i)=0$, $\bar{g}(r_i)=-v^\ast_i(r_i)$ and $h(\theta_{-i})=v^\ast_i(\emptyset)$, for any two invitation strategies $r_i^1\subseteq r_i^2$, since $\tilde{g}(r_i)=0$, then $\tilde{g}(r_i^1)=\tilde{g}(r_i^2)=0$, i.e., $r_i =\arg\min_{r_i^\prime \subseteq r_i}\tilde{g}_i(r_i^\prime)$; $\bar{g}(r_i^1)=-v^\ast_i(r_i^1) \geq v^\ast_i(r_i^2) = \bar{g}(r_i^2)$ (according to \Cref{critical_bid_monotonicity_competitive}), i.e., $r_i = \arg\min_{r_i^\prime \subseteq r_i}\bar{g}_i(r_i^\prime)$.

For IP-MON allocations, set $\tilde{g}(r_i)=v^\ast_i(r_i)$, $\bar{p}(r_i)=0$, and $h(\theta_{-i})=0$. for any two invitation strategies $r_i^1\subseteq r_i^2$, $\tilde{g}(r_i^1)=v^\ast_i(r_i^1) \geq v^\ast_i(r_i^2)=\tilde{g}(r_i^2)$ (according to \Cref{critial_bid_IP_monotonicity}). Thus, $r_i =\arg\min_{r_i^\prime \subseteq r_i}\tilde{g}(r_i^\prime)$; $\bar{g}(r_i^1)=\bar{g}(r_i^2)=0$, then $r_i =\arg\min_{r_i^\prime \subseteq r_i}\bar{g}(r_i^\prime)$. 
\end{proof}

\subsection{Proof of \Cref{EFF_WBB_2_MONO}}
\begin{proof}
    In \Cref{VCG_RM_properties}, we have shown the efficient allocation rule satisfies ID-MON, now we show the efficient allocation is not IP-MON. 
    
    Consider the following simple example: seller $s$ has one unit good available, the market is $G=(N\cup\{s\}, E)$ where $N=\{A,B\}$ and $E=\{(s,A), (A,B)\}$. For $A$, $v_A=1,r_A=\{B\}$; For $B$, $v_B=2, r_B=\emptyset$. In this case, to be efficient, the unique item will be allocated to $B$ and $A$ gets nothing, i.e., $f_A((v_A,r_A),\theta_{-A})=0$. However, if $A$ misreport $r_A^\prime=\emptyset$, bidder $B$ cannot enter the market and the efficient allocation will allocate the item to $A$, i.e., $f_A((v_A,r_A^\prime),\theta_{-A})=1 > f_A((v_A,r_A),\theta_{-A})$. Note that according to the definition of IP-MON, $(v_A,r_A^\prime) \succeq_{\mathcal{P}} (v_A,r_A)$, which implies $f_A((v_A,r_A),\theta_{-A})\geq f_A((v_A,r_A^\prime),\theta_{-A})$. This leads to a contradiction. Therefore, the efficient allocation cannot be IP-MON. 

    For any $\mathcal{M}=(f,p)$ where $f$ satisfies IP-MON and $p$ is revenue-maximizing payment in \Cref{opt_payment_IP_monotonicity}.  $\mathcal{M}$ satisfies WBB because only winning bidders pay some money to the seller while all the other losing bidders always pay zero, i.e., $\Rev^{\mathcal{M}}=\sum_{i:f_i=1}\tilde{p}_i(r_i)\geq 0$. For ID-MON allocation with revenue-maximizing payment, it may not be WBB. The most direct example is that the extended VCG mechanism in single-item network auction is efficient, IR, SP, but not WBB \cite{LHZ20a,LHG+22a}. 
\end{proof}

\subsection{Proof of \Cref{ID_IP_MON_computational_tractable}}
\begin{proof}
According to the revenue-maximizing payment scheme $p^\ast$, for each bidder $i$, her payment $p^\ast_i$ is decided by two critical bids $v_i^\ast(\emptyset)$ and $v_i^\ast(r_i)$, thus given an ID-MON or IP-MON monotone allocation $f$, for each bidder $i$, we can run $f$ in profile $\theta=((v_i, \emptyset), \theta_{-i})$ and binary search for the critical bid $v_i^\ast(\emptyset)$ such that bidder $i$ becomes a winner in $\theta=((v_i^\ast(\emptyset), \emptyset), \theta_{-i})$. The same binary search procedures can be done for computing $v_i^\ast(r_i)$. The critical bid $v_i^\ast(\emptyset)$ and $v_i^\ast(r_i)$ should be computed for every bidder $i\in N$ while the value interval for binary search is $[0, \max_{i\in N}v_i]$. Therefore, the revenue-maximizing payment can be computed in $O(N\cdot T\log(\max_{i\in N}v_i))$.
\end{proof}

\subsection{Example for comparing existing strategyproof mechanisms}

\begin{figure}[!htbp]
    \centering
    \scalebox{1}{
\begin{tikzpicture}[global scale=0.99, box/.style={circle, draw}]
            \node[box,fill=gray,very thin](s) at(0,-0.5){{\LARGE$s$}};
            \node[box,fill={rgb:red,1;green,180;blue,80}](A) at(1.2,-1){$A$};
            \node[] at (1.7, -.7){$3$};
            \node[box,fill=white](H) at(2.4,-1){$H$};
            \node[] at (2.9, -.7){$2$};
            \node[box,fill=white](I) at(4.8,-1){$I$};
            \node[] at (5.3, -.7){$4$};
            
            \node[box,fill={rgb:red,1;green,180;blue,80}](B) at(1.2,0){$B$};
            \node[] at (1.7, 0.3){$1$};
            \node[box,fill=white](C) at(2.4,0){$C$};
            \node[] at (2.9, 0.3){$2$};
            \node[box,fill=white](D) at(3.6,0){$D$};
            \node[] at (4.1, 0.3){$100$};
            \node[box,fill=white](E) at(4.8,0){$E$};
            \node[] at (5.3, 0.3){$5$};
            \draw[-,  line width=.8pt] (s) --(A);
            \draw[-,  line width=.8pt] (A) --(H);
            \draw[-,  line width=.8pt]  (s) --(B);
            \draw[-,  line width=.8pt]  (B) --(C);
            \draw[-,  line width=.8pt]  (C) --(D);
            \draw[-,  line width=.8pt]  (D) --(E);
            \draw[-,  line width=.8pt]  (D) --(I);
        \end{tikzpicture}}
    \caption{$3$-unit network auction with unit demand bidders.}
    \label{example-comparsion}
\end{figure}
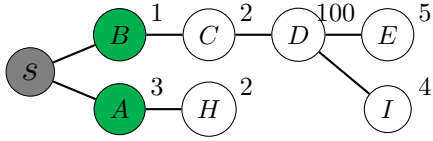

The example in \Cref{example-comparsion} features seven potential bidders, with only bidders A and B having direct access to the auction message from the seller. The seller is offering three homogeneous goods. \Cref{mechanism_comparsion} presents detailed results from running five existing strategyproof mechanisms. In \Cref{mechanism_comparsion}, in the payment column, all the negative numbers are the money that the seller should pay to the bidders to incentivize information diffusion. In the $\mathcal{M}$ column, we highlight two novel mechanisms proposed in this paper: VCG-RM and DNA-MU-R. 

\begin{table*}[!htbp]
\centering
\caption{Allocation and payment results for \Cref{example-comparsion}. }\label{mechanism_comparsion}
\scalebox{1}{
\begin{tabular}{ccccl}
\toprule
$\mathcal{M}$  & $\SW $  & $\Rev$      &  Winner      & Payment  \\ 
\midrule
VCG & $\mathbf{109}$ & $-203$ & $\{D,E,I\}$ & $A(0),B(-104), C(-103),D(-2),E(3),H(0),I(3)$ \\
\midrule
\textbf{VCG-RM} & $\mathbf{109}$ & $1$ & $\{D,E,I\}$ & $A(0),B(-4), C(-3), D(2),E(3),H(0),I(3)$  \\
\midrule
\textbf{DNA-MU-R} & {$103$} & {$\mathbf{4}$} & {$\{B,C,D\}$} & $A(0),B(0), C(1), D(3),E(0),H(0),I(0)$  \\
\midrule 
{LDM-Tree} & {$6$} & {$2$} & {$\{A,B,C\}$} & $A(0),B(0), C(2), D(0),E(0),H(0),I(0)$  \\
\midrule 
{MUDAN} & {$6$} & {$2$} & {$\{A,B,C\}$} & $A(0),B(0), C(2), D(0), E(0),H(0),I(0)$  \\
\bottomrule 
\end{tabular}}  
\end{table*}

\begin{table*}[!htbp]
\centering
\caption{Classification of Existing Strategyproof Network Auctions (Mechanisms Proposed in This Paper Highlighted in Bold)}
\label{classification_mechanism_monotonicity_table}
\begin{tabular}{@{}p{3cm}p{3.5cm}p{3.5cm}p{4cm}@{}}
\toprule
& \textbf{ID-MON}  & \textbf{IP-MON}  & \textbf{Degenerated} \\ 
\multicolumn{1}{l}{Settings}    & \multicolumn{1}{l}{}   & \multicolumn{1}{l}{}  & \multicolumn{1}{l}{}   \\ \midrule

\multicolumn{1}{l}{\textbf{Single-item}}     & \multicolumn{1}{l}{\begin{tabular}[c]{@{}l@{}}IDM \citep{LHZ+17a} \end{tabular}} &  & \multicolumn{1}{l}{DNA-MU \citep{KBT+20a}} \\ \midrule

\multicolumn{1}{l}{\textbf{Unit-demand}} & \multicolumn{1}{l}{\begin{tabular}[c]{@{}l@{}}VCG \citep{VICK61a}\\\textbf{VCG-RM} \\ LDM \citep{LLZ23a} \end{tabular}} & \multicolumn{1}{l}{\begin{tabular}[c]{@{}l@{}}\textbf{DNA-MU-R} [Alg \ref{FIX_DNA_MU}]\\ MUDAN \citep{FZL+23a}\end{tabular}} &Diff-CRA-HM \citep{GHX+23a} \\ \midrule

\multicolumn{1}{l}{\textbf{With Budget}} & &\multicolumn{1}{l}{\begin{tabular}[c]{@{}l@{}} SNCA \citep{XSK22a} \end{tabular}} & \multicolumn{1}{l}{\begin{tabular}[c]{@{}l@{}} BDM-H/G \citep{LWL+21a} \end{tabular}} \\ \midrule

\multicolumn{1}{l}{\textbf{Double Auction}} & &\multicolumn{1}{l}{\begin{tabular}[c]{@{}l@{}} DTR \citep{LCZ24a} \end{tabular}} &  \\ \midrule
\multicolumn{1}{l}{\textbf{Single-minded}} & \textbf{Net-$\sqrt{k}$-APM} &\multicolumn{1}{l}{\begin{tabular}[c]{@{}l@{}} \\ 
 \end{tabular}} & \textbf{NSA} \\
\bottomrule
\end{tabular}
\end{table*}

The VCG-RM mechanism is efficient, achieving optimal social welfare while significantly mitigating the deficit issues inherent to the classic VCG mechanism (with a deficit at $203$). Additionally, VCG-RM attains the highest possible revenue under an efficient allocation. The DNA-MU-R mechanism, on the other hand, substantially outperforms the LDM-Tree and MUDAN mechanisms in terms of social welfare and yields the highest revenue in this particular example. Notably, VCG-RM mechanism is built upon the concept of ID-MON allocation with revenue-maximizing payments (\Cref{opt_payment_id_monotonicity}), while DNA-MU-R mechanism is based on IP-MON allocation with revenue-maximizing payments (\Cref{opt_payment_IP_monotonicity}). This example underscores the effectiveness of our newly proposed ID-MON and IP-MON auction design principles.

\subsection{A Classification of Existing SP Network Auction Mechanisms}\label{classification_existing_sp_mechanisms}

We classify all existing strategyproof network auction mechanisms into three categories based on the monotonicity of their allocation rules: ID-MON, IP-MON, and Degenerated, as shown in \Cref{classification_mechanism_monotonicity_table}. The mechanisms proposed in this paper are highlighted in bold. The VCG-RM mechanism is introduced in Section 4.2. The DNA-MU-R mechanism is presented as a corrected version of DNA-MU, restoring its strategyproofness. In Section 5, we propose Net-$\sqrt{k}$-APM and NSA, which can be seen as two distinct extensions of the classical $\sqrt{k}$-approximation allocation rule.

\section{Omitted Contents from Section 5}\label{omitted_proof_section5}

\subsection{Proof of \Cref{sqrt_k_approx_mechanism}}
\begin{proof}
    The Net-$\sqrt{k}$-APM mechanism is $\sqrt{k}$-efficient since it adopts the same allocation rule in classic single-minded auction setting applied to all the bidders in $N$. To show the Net-$\sqrt{k}$-APM mechanism satisfies IR and SP. it suffices to show that its allocation rule satisfies ID-MON. Consider any bidder $i\in N$.
    
    \textbf{Valuation}. Fix all other bidders' profile $\theta_{-i}$ and $i$'s invitation strategy $r_i$. Suppose bidder $i$ is a winner under $v_i$. It follows that there is no bundle-conflicting bidder has higher rank than $i$ for $\frac{v_i}{\sqrt{|S_i^\ast|}}$. Now if $i$ increases her bid to $v_i^\prime$, it holds that $\frac{v_i^\prime}{\sqrt{|S_i^\ast|}} \geq \frac{v_i}{\sqrt{|S_i^\ast|}}$, and bidder $i$ will maintain or improve her rank, and no additional bundle-conflicting bidders will outrank her. Therefore, bidder $i$ remains eligible to win, satisfying monotonicity in valuation.
    
    \textbf{Invitation}: Fix all other bidders' profile $\theta_{-i}$ and $i$'s bid $v_i$. Suppose $i$ wins under invitation strategy $r_i$, i.e., $f_i(v_i, r_i) = 1$. Consider any subset $r_i' \subseteq r_i$. The only change from $r_i$ to $r_i'$ is the exclusion of some invited bidders (i.e., $r_i \setminus r_i'$), which are no longer considered in the ranking based on $\frac{v_i}{\sqrt{|S_i^\ast|}}$. The only possible way $i$ could lose is if some bundle-conflicting bidder with a higher ranking displaces her. However, reducing the invitation set can only reduce the number of such potentially conflicting bidders. Hence, if $i$ wins with $(v_i, r_i)$, she must also win with $(v_i, r_i')$. This confirms monotonicity in invitation.

    Monotonicities over both valuation and invitation imply that the allocation rule satisfies ID-MON. According to \Cref{opt_payment_id_monotonicity}, by applying the corresponding payment rule in \Cref{opt_payment_id_monotonicity}, the Net-$\sqrt{k}$-APM mechanism satisfies IR and SP.

    Regarding WBB, consider a scenario in which there is only one item for sale and each bidder's favorite bundle is the unique item. In this case, it degenerates to the classic single-item auction and the allocation rule of Net-$\sqrt{k}$-APM mechanism becomes EF. However, by the impossibility results in single-item network auction, no mechanism can simultaneously satisfy EF, IR, SP, and WBB (Theorem 2 in \cite{LHG+22a}). So the Net-$\sqrt{k}$-APM mechanism necessarily violates WBB in this scenario.
\end{proof}

\subsection{Omitted Proof of \Cref{single_minded_IP_MON}}
\begin{proof}
    To show that the NSA mechanism satisfies IR, SP, and WBB. From \Cref{opt_payment_IP_monotonicity}, it suffices to prove the allocation rule in \Cref{mechanism_for_single_minded_bidders} satisfies IP-MON. We next prove the monotonicity regarding the valuation and invitation dimension, respectively. Consider any bidder $i\in N$.

    \textbf{Valuation}. Fix the profile $\theta_{-i}$ of all other bidders and $i$'s invitation strategy $r_i$. To prove the value-monotonicity, we aim to show that for any $v_i' \geq v_i$, if $f(v_i, r_i)=1$, it must hold that $f(v_i',r_i)=1$. Consider agent $i$ misreporting her bid from $v_i$ to a higher value $v_i'$. The first observation is that $i$ is still the agent who maximizes the value of $\frac{v_i}{\sqrt{|S_i^*|}}$ in $N_{-T_i}$ since $\frac{v_i'}{\sqrt{|S_i^*|}} \geq \frac{v_i}{\sqrt{|S_i^*|}}$. Hence, the only condition we need check is whether it still holds that $S_i^*\cap (\bigcup_{j\in W} S_j^*)=\emptyset$. Let $W'$ denote the new winner set after $i$'s misreport. We observe that $W' \subseteq W$ as the only possible effect of this misreport is that some other agent $j$ with higher priority is displaced from the winner set under $v_i'$. So we have $S_i^*\cap (\bigcup_{j\in W'} S_j^*)=\emptyset$ as well. Therefore, agent $i$ remains a winner after misreporting to any $v_i' \geq v_i$.
    
    \textbf{Invitation}. Fix the profile $\theta_{-i}$ of all other bidders and agent~$i$'s bid strategy $v_i$. Consider two invitation strategies $r_i'$ and $r_i$ such that $r_i' \subseteq r_i$. We aim to show that if $f(v_i, r_i') = 1$, then it must also hold that $f(v_i, r_i) = 1$. 
    
    Notice that agent $i$'s winning condition depends on two factors: (i). whether $\frac{v_i}{\sqrt{|S_i^*|}}$ is the highest among all of the agents in $N_{-T_i}$, and (ii) whether $i$'s favorite bundle is disjoint from those of previously selected winners in $W$ at the time of eligibility checking. The first condition—$\frac{v_i}{\sqrt{|S_i^|}} \geq \frac{v_j}{\sqrt{|S_j^|}}$ for all $j \in N_{-T_i}$ is independent of the invitation strategy. Hence, we focus on how $i$'s invitation strategy affects the resulting winner set $W$. 
    
    Since $i$ is a winner under $r_i'$, we have $\frac{v_i}{\sqrt{|S_i^*|}} \geq \frac{v_j}{\sqrt{|S_j^*|}}$ for any $j\in N_{-T_i}'$ (here $N_{-T_i}'$ is the set under $r_i'$). For any agent $j \in N_{-T_i}'$ who has higher priority than $i$ in the processing order $\mathcal{O}$, and who is selected as a winner, it must be the case that $\frac{v_j}{\sqrt{|S_j^|}} = \frac{v_i}{\sqrt{|S_i^|}}$—otherwise, $i$ would not be eligible to win.
    
    Recall that agent $i$ is a winner under $r_i'$. We consider how the inclusion of additional agents in $r_i$ (i.e., $r_i \setminus r_i'$) may affect the outcome. We distinguish two sub-cases:
    
    \textbf{Sub-case (1).} For every agent $q\in r_i\setminus r_i'$, if $\frac{v_q}{\sqrt{|S_q^*|}} \leq \frac{v_i}{\sqrt{|S_i^*|}}$, then when $i$ reports $r_i$, the allocation remains unchanged and $i$ remains a winner; 
    
    \textbf{Sub-case (2).} Suppose there exists some $q \in r_i \setminus r_i'$ such that $\frac{v_q}{\sqrt{|S_q^|}} > \frac{v_i}{\sqrt{|S_i^|}}$. When agent $i$ reports $r_i$, any winner $j\in N_{-T_i}'$ with higher priority than $i$ will loss because of the existence of $q$. 
    
    This implies that when $i$'s invitation strategy changes from $r_i'$ to $r_i$, it could only be either the allocation remains unchanged or there are fewer winners in the winning set. In both cases, agent $i$ remains winning, implying that the allocation rule satisfies IP-MON.
    
    Since the allocation rule of the NSA mechanism has been shown to satisfy IP-MON, combining it with the payment scheme from \Cref{opt_payment_IP_monotonicity} ensures that the NSA mechanism satisfies IR and SP. Moreover, it trivially satisfies WBB as the seller never makes payments to the bidders under the NSA mechanism.
\end{proof}

\subsection{Omitted Discussion Regarding IP-MON in Single-minded Network Auction}
In this section, we present a brief discussion on the IP-MON allocation rule in single-minded network auctions, emphasizing the challenges in identifying a desirable IP-MON-compliant allocation. The mechanism composed of \Cref{mechanism_for_single_minded_bidders} and the revenue-maximizing payment scheme in \Cref{opt_payment_IP_monotonicity} turns out to be degenerated. In addition to the allocation rule described in \Cref{mechanism_for_single_minded_bidders}, we also examine several alternative allocation rules, which, unfortunately, fail to satisfy IP-MON.

The first one tries to extend the \Cref{sqrt_k_approximation} allocation into the networked scenario. 
\begin{algorithm}[!htbp]
\caption{Exploratory Allocation Rule (I)}
\label{alg:LOS_extension_fails_IP_MON}
\begin{algorithmic}[1]
\REQUIRE $G=(N\cup\{s\}, E)$, $\theta$, $\mathcal{K}$;
\ENSURE Allocation $f$;
\STATE Initialize winner set $W$;
\STATE $\mathcal{O}\leftarrow \texttt{BFS}(G,s)$; Create the IDT $T$;
\FOR{Bidder $i$ in $\mathcal{O}$}
\STATE $N_{-T_i}\leftarrow N \setminus T_i \cup \{i\}$;\\
\STATE $\theta_{-T_i}\leftarrow \{(v_j,S_j^\ast)\}_{j\in N_{-T_i}}$, $\bar{W}_i \leftarrow W$;\\
\STATE $\bar{W}_i \leftarrow $ \texttt{\Cref{sqrt_k_approximation}}($\theta_{-T_i}$, $\bar{W}_i$);\\
\IF{$i$ in $\bar{W}_i$}
\STATE Update $W\leftarrow W \cup\{i\}$;
\ENDIF
\ENDFOR
\STATE Return $f$ which gives $S_i^\ast$ to $i$ if and only if $i\in W$.
\end{algorithmic}
\end{algorithm}

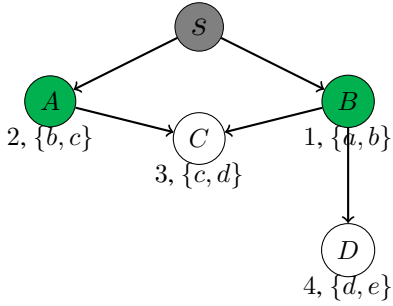
\begin{figure}[!htbp]
    \centering
    \scalebox{1}{
\begin{tikzpicture}[global scale=0.99, box/.style={circle, draw}]
            \node[box,fill=gray,very thin](s) at(0,2){{\LARGE$s$}};
            \node[box,fill={rgb:red,1;green,180;blue,80}](A) at(-2,1){$A$};
            \node[] at (-2, 0.5){$2$, $\{b,c\}$};
            \node[box,fill={rgb:red,1;green,180;blue,80}](B) at(2,1){$B$};
            \node[] at (2, 0.5){$1$, $\{a,b\}$}; \node[box](C) at(0,0.5){$C$};
            \node[] at (0,0){$3$, $\{c,d\}$}; 
            \node[box](D) at(2,-1){$D$};
            \node[] at (2, -1.5){$4$, $\{d,e\}$}; 
            \draw[->,  line width=.8pt] (s) --(A);
            \draw[->,  line width=.8pt]  (s) --(B);
            \draw[->,  line width=.8pt]  (A) --(C);
            \draw[->,  line width=.8pt]  (B) --(C);
            \draw[->,  line width=.8pt]  (B) --(D);
        \end{tikzpicture}}
    \caption{Counterexample in which Exploratory Allocation Rule (I) fails IP-MON}
\label{fig:IP_MON_fails_example_1} 
\end{figure}

However, this algorithm fails to satisfy IP-MON, as demonstrated by the following counterexample. Consider an instance with four agents, ${A, B, C, D}$, and five heterogeneous items, ${a, b, c, d, e}$, available for sale. Each agent's preferred bundle and corresponding valuation are detailed in \Cref{fig:IP_MON_fails_example_1}.

Consider the agent ordering $\mathcal{O} = (A, B, C, D)$. We show that agent $B$ violates IP-MON under \Cref{alg:LOS_extension_fails_IP_MON}. Specifically, for agent $B$, we have $f(v_B, {D}) = 0$ while $f(v_B, \emptyset) = 1$, contradicting the monotonicity requirement.

To see this, suppose $r_B = {D}$. When processing agent $A$, we run \Cref{sqrt_k_approximation} on the set ${A, B, C, D}$ and obtain $A$ and $D$ as winners. Since $A$ is selected into $W$, agent $B$ becomes ineligible due to the overlap $S_B^* \cap S_A^* = {b} \neq \emptyset$.

In contrast, when $r_B = \emptyset$, agent $A$ considers only ${A, B, C}$. Running \Cref{sqrt_k_approximation} on this subset yields $B$ and $C$ as winners. As a result, $A$ is not included in $W$, and the mechanism proceeds to agent $B$. At this point, running \Cref{sqrt_k_approximation} again on ${A, B, C}$ selects $B$ as a winner, and thus $B$ is added to $W$.

This discrepancy in outcomes based on $r_B$ demonstrates that \Cref{alg:LOS_extension_fails_IP_MON} fails to satisfy IP-MON.

We propose an alternative allocation rule, inspired by DNA-MU and \Cref{sqrt_k_approximation}, which is formally defined in \Cref{alg:exploratory_allocation_II}.

\begin{algorithm}[!htbp]
\caption{Exploratory Allocation Rule (II)}
\label{alg:exploratory_allocation_II}
\begin{algorithmic}[1]
\REQUIRE $G=(N\cup\{s\}, E)$, $\theta$, $\mathcal{K}$;
\ENSURE Allocation $f$;
\STATE Initialize winner set $W$ and integer parameter $k$;
\STATE $\mathcal{O}\leftarrow \texttt{BFS}(G,s)$; Create the IDT $T$;
\FOR{Bidder $i$ in $\mathcal{O}$}
\STATE $N_{-T_i}\leftarrow \left(N \setminus T_i\right) \cup \{i\}$;
\IF{$i$ ranks top-$k$ among $N_{-T_i}$ w.r.t. $\frac{v_i}{\sqrt{|S_i^*|}}$ and $S_i^\ast \cap (\bigcup_{j\in W}S^\ast_j)=\emptyset$}
\STATE Update $W\leftarrow W \cup\{i\}$;
\ENDIF
\ENDFOR
\STATE Return $f$ which gives $S_i^\ast$ to $i$ if and only if $i\in W$.
\end{algorithmic}
\end{algorithm}

\begin{figure}[!htbp]
    \centering
    \scalebox{1}{
\begin{tikzpicture}[global scale=0.99, box/.style={circle, draw}]
            \node[box,fill=gray,very thin](s) at(0,2){{\LARGE$s$}};
            \node[box,fill={rgb:red,1;green,180;blue,80}](A) at(-2,1){$B$};
            \node[] at (-2, 0.5){$3$, $\{a,b\}$};
            \node[box,fill={rgb:red,1;green,180;blue,80}](B) at(2,1){$C$};
            \node[] at (2, 0.5){$3$, $\{c,d\}$}; \node[box,fill={rgb:red,1;green,180;blue,80}](C) at(0,0){$A$};
            \node[] at (0,-.6){$3$, $\{b,c\}$}; 
            \node[box](D) at(2,-1){$D$};
            \node[] at (2, -1.6){$4$, $\{e\}$}; 
            \node[box](E) at(-2,-1){$E$};
            \node[] at (-2, -1.6){$4$, $\{e\}$}; 
            \draw[->,  line width=.8pt] (s) --(A);
            \draw[->,  line width=.8pt]  (s) --(B);
            \draw[->,  line width=.8pt]  (s) --(C);
            \draw[->,  line width=.8pt]  (A) --(E);
            \draw[->,  line width=.8pt]  (B) --(D);
        \end{tikzpicture}}
    \caption{Counterexample in which Exploratory Allocation Rule (II) fails IP-MON}
\label{fig:IP_MON_fails_example_2} 
\end{figure}
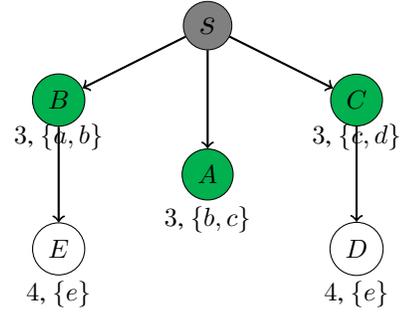

\Cref{alg:exploratory_allocation_II} also fails to satisfy IP-MON, as demonstrated by the counterexample in \Cref{fig:IP_MON_fails_example_2}. In this example, there are five agents, ${A, B, C, D, E}$, and five heterogeneous items, ${a, b, c, d, e}$.

Let the parameter $k = 2$, and consider the agent ordering $\mathcal{O} = (A, B, C, D, E)$. We show that \Cref{alg:exploratory_allocation_II} violates IP-MON by showing that, for agent $B$, we have $f(v_B, {E}) = 0$ while $f(v_B, \emptyset) = 1$.

When $r_B = {E}$, we begin with agent $A$. Agent $A$ cannot be selected as a winner, as both $D$ and $E$ have higher scores than $A$ with respect to the ranking criterion $\frac{v_i}{\sqrt{|S_i^*|}}$. The mechanism then proceeds to agent $B$, who is eligible to win since, in the set $N_{-T_B}$, agents $B$ and $D$ are ranked among the top two. As a result, $B$ is selected as a winner.
However, when $r_B = \emptyset$, agent $A$ changes status from a loser to a winner. This prevents agent $B$ from being selected due to a bundle conflict, as $S_B^* \cap S_A^* = {b} \neq \emptyset$. In this case, $B$ loses her winning status when misreporting $r_B = \emptyset$, violating the monotonicity condition.It implies that the allocation rule in \Cref{alg:exploratory_allocation_II} fails IP-MON.

In essence, the primary reason why \Cref{alg:LOS_extension_fails_IP_MON} and \Cref{alg:exploratory_allocation_II} fail to satisfy IP-MON lies in the dual effect of invitations on market dynamics. For any agent $i$, inviting additional participants can have conflicting impacts on another agent $j$: the newcomers may introduce greater competition, thereby reducing $j$'s chances of winning, or conversely, they may alter the allocation landscape in a way that increases $j$’s likelihood of being selected. The latter effect arises from the intricate combinatorial structure of bundle conflicts among agents, which can inadvertently benefit some participants. This interplay highlights the inherent complexity in designing allocation rules that satisfy IP-MON in single-minded network auctions.

\end{document}